\documentclass[fleqn,usenatbib]{mnras}

\usepackage{newtxtext,newtxmath}

\usepackage[T1]{fontenc}

\DeclareRobustCommand{\VAN}[3]{#2}
\let\VANthebibliography\thebibliography
\def\thebibliography{\DeclareRobustCommand{\VAN}[3]{##3}\VANthebibliography}

\usepackage{graphicx}	
\usepackage{amsmath}	
\usepackage{comment}
\usepackage{siunitx}
\usepackage{caption}
\usepackage{subcaption}
\usepackage{multirow}
\usepackage{rotating}
\usepackage{booktabs}
\usepackage{enumitem}
\usepackage{arydshln}
\usepackage{pdflscape}

\newcommand{\kms}{km\,s$^{-1}$} 
\newcommand{\pcmm}{\,cm$^{-2}$}

\newcommand{\pc}{\,pc$^{-2}$}

\newcommand{\fw}{$f_\mathrm{W}$}
\newcommand{\lclump}{$L_\mathrm{clump}$}

\newcommand{\mclump}{$M_\mathrm{clump}$}

\newcommand{\mgmc}{$M_\mathrm{GMC}$}

\title[Kinematic distances of WISE \ion{H}{ii} regions]{Determination of kinematic distances of WISE \& SMGPS \ion{H}{ii} regions in the Galactic plane using SEDIGISM cloud association}

\author[M. O. Langa et al.]{
Moses O. Langa,$^{1,2}$\thanks{E-mail: langamoses1@gmail.com}
Mark A. Thompson,$^{2}$
Andrew J. Rigby,$^{2}$
Gwenllian M. Williams,$^{3}$
Mubela Mutale,$^{2}$
\newauthor
Cristobal Bordiu,$^{4}$
Simone Riggi,$^{4}$
Willice O. Obonyo,$^{1,5}$
James O. Chibueze,$^{5,6}$
and Paul Baki$^{1}$
\\
$^{1}$Department of Physics, Earth and Environmental Sciences, Technical University of Kenya, Nairobi, P.O. Box 52428-00200, Kenya\\
$^{2}$School of Physics and Astronomy, University of Leeds, Leeds LS2 9JT, UK\\
$^{3}$Department of Physics, Aberystwyth University, Ceredigion, Cymru, SY23 3BZ, UK\\
$^{4}$ INAF-Osservatorio Astrofisico di Catania, Via Santa Sofia 78, 95123 Catania, Italy\\
$^{5}$Centre for Astrophysics and Space Sciences, University of South Africa, Cnr
Christian de Wet Rd and Pioneer Avenue, Florida Park, 1709, Roodepoort,
South Africa\\
$^{6}$Department of Physics and Astronomy, Faculty of Physical Sciences, University of Nigeria, Carver Building, 1 University Road, Nsukka 410001, Nigeria\\
}

\date{Accepted 2026 July 02. Received 2026 July 02; in original form 2026 February 23}

\pubyear{\the\year{}}

\begin{document}
\label{firstpage}
\pagerange{\pageref{firstpage}--\pageref{lastpage}}
\maketitle

\begin{abstract}
One of the fundamental requirements for studying and understanding Galactic structure and massive star formation is accurate distances to \ion{H}{ii} regions. However, most distance assignments are hampered by kinematic distance ambiguity (KDA), sparse parallax measurements, and the large number of radio continuum sources lacking velocity and distance measurements. We present a kinematic distance determination method via cloud association, linking \ion{H}{ii} regions from the  Wide-field Infrared Survey Explorer (WISE) catalogue and the South African Radio Astronomy Observatory (SARAO) MeerKAT Galactic Plane Survey (SMGPS) with molecular clouds from the Structure, Excitation and Dynamics of the Inner Galactic Interstellar Medium (SEDIGISM) $^{13}$CO (2--1) survey. The associations are established through spatial overlap and velocity coherence, and the molecular cloud velocity and distance are then assigned to the associated \ion{H}{ii} region. The method yields 741 \ion{H}{ii} regions with adopted CO-based systemic velocities, of which 640 have reliable kinematic distances based on the SEDIGISM distance reliability criteria. We validate the method using 329 \ion{H}{ii} regions with independent radio recombination line (RRL) velocities, finding excellent agreement with a median absolute velocity difference of 3.46\kms. Our analysis resolves ambiguous velocities for 40 \ion{H}{ii} regions with multiple WISE RRL velocity measurements. Compared to the unassociated clouds, the associated molecular clouds exhibit significantly higher masses, gas surface densities, linewidths, star formation efficiencies and dense gas fractions, and slightly lower virial parameters. This work provides a large, homogeneously-derived catalogue of \ion{H}{ii} region distances and establishes a framework for further studying massive star formation and Galactic structure in general.
\end{abstract}




\section{Introduction}
\label{sec: Intro}
Massive star–forming sites embedded in the Galactic disc \citep{2013MNRAS...Urquhart_a,2014MNRAS...Urquhart} play a crucial and central role both as tracers of large-scale Galactic structure and as drivers of the evolution of the interstellar medium through stellar feedback processes. Ionised hydrogen regions (\ion{H}{ii} regions) mark sites of recent massive star formation and are therefore fundamental to studies of Galactic spiral structure \citep[][]{2014ApJ...Reid, 2019ApJ...Reid}, cluster formation \citep{2003ARA&A..Lada}, and possibly triggered star formation scenarios \citep[e.g.][]{2005A&A...Deharveng, 2010A&A...Zavagno, 2012ApJ...Kendrew}. Giant Molecular Clouds (GMCs) represent the cold gas reservoirs from which massive stars are born \citep{2014prpl.Dobbs}, and the spatial or kinematic association of \ion{H}{ii} regions with the molecular clouds provides a powerful and significant means to study the birth, environment and impact of massive stars \citep{2012AAS...Bressert}.

\ion{H}{ii} regions are among the most accessible tracers of massive‐star formation in the Galaxy: their free‐free emission or radio continuum \citep[][]{1967ApJ...Mezger, 2009tra.....Wilson}, recombination‐line emission and mid-infrared (MIR) signatures \citep[][]{2005pcim.book...Tielens, 2008ApJ...Watson} allow identification across wide Galactic longitudes. For example, the Wide-field Infrared Survey Explorer (WISE) Catalogue of Galactic \ion{H}{ii} Regions \citep[][]{2014ApJS..Anderson, 2015ApJS..Anderson, 2018ApJS..Anderson} compiles over 8000 \ion{H}{ii} regions and candidates throughout the Galactic plane, including $\sim$1500 with measured radio recombination lines and/or distances. These \ion{H}{ii} regions in the WISE catalogue can easily be identified by the SARAO MeerKAT 1.3 GHz Galactic Plane Survey \citep[SMGPS;][]{2024MNRAS..Goedhart} which provides a well-selected and uniform sample of \ion{H}{ii} regions in an extended source catalogue \citep[][]{2025A&A...Bordiu}, containing 3326 extended \ion{H}{ii} regions, alongside additional extended Galactic sources not included in this count. However, the utility of \ion{H}{ii} regions as tracers of Galactic structure, spiral arms, and massive star formation is critically dependent on obtaining reliable distances: uncertainties in heliocentric or Galactocentric distance propagate into uncertainties in derived physical parameters such as size, luminosity, ionizing photon flux \citep{2010ApJ...Murray} and mass of associated gas \citep[e.g.][]{2017A&A..Kauffmann, 2017A&A..Kauffmann_2}, and into their placement within spiral‐arm frameworks \citep[e.g.][]{2014ApJ...Reid, 2019ApJ...Reid}.

Many \ion{H}{ii} region distances are derived from kinematics (radial velocities plus a Galactic rotation model). But kinematic distances suffer from several well‐known issues: departures from purely circular rotation (streaming motions, spiral‐arm shocks), the kinematic distance ambiguity (KDA) for inner‐Galaxy sightlines (near and far--distance ambiguity), and uncertain rotation‐curve parameters \citep[][]{2012PASJ..Sakai, 2015PASJ...Sakai, 2019ApJ...Reid}. For instance, using then existing \ion{H}{i} and $^{13}$CO sky surveys, \cite{2009ApJ...Anderson} resolved the KDA for 266 inner Galaxy \ion{H}{ii} regions from a sample of 291 and found that features such as the far/near ratio vary significantly by morphological class (ultra‐compact versus diffuse). More recently, parallax‐ and photometric‐based distances (e.g., via Gaia or masers) for \ion{H}{ii} regions remain relatively sparse, meaning that for most Galactic \ion{H}{ii} regions the reliance on kinematic distances remains. For example, a recent study by \cite{2025A&A...Shen} compiled 459 \ion{H}{ii} regions with Gaia‐derived distances and showed corrections compared to earlier methods. In this context, any approach that can improve the robustness, sample size or reliability of \ion{H}{ii} region distances is of significant value for Galactic structure and star‐formation studies.

Connecting \ion{H}{ii} regions to their natal or adjacent molecular material is crucial for understanding their context \citep[][]{2014prpl.Dobbs, 2018NatAs...Motte_a, 2018ARA&A..Motte_b, 2022A&A...Neralwar, 2022A&A...Neralwar_b}. Surveys such as the Structure, Excitation and Dynamics of the Inner Galactic InterStellar Medium \citep[SEDIGISM;][]{2021MNRAS..Schuller} survey have extracted over 10000 molecular clouds in the inner Galaxy using $^{13}$CO (2--1) data. For example, \cite{2021MNRAS.Duarte} report a SEDIGISM catalogue of 10663 molecular clouds, of which $\sim$10300 have assigned distances (which were later updated by \cite{2022A&A...Colombo}) and derived physical properties. The SEDIGISM approach uses a Galactic rotation model \citep[e.g.,][]{2016ApJ...Reid} and a combination of KDA resolution methods such as maser associations, infrared dark cloud (IRDC), \ion{H}{i} self‐absorption (HISA), extinction, size–linewidth relations to resolve the KDA, and also updated by \cite{2022A&A...Colombo} using spiral arm association for some clouds whose distance solution remained uncertain with the other methods, building on methodologies developed in other large-scale surveys \citep[e.g.,][for the APEX Telescope Large Area Survey of the Galaxy (ATLASGAL)]{2012A&A...Wienen}. Because molecular clouds frequently host \ion{H}{ii} regions via ongoing massive‐star formation \citep[][]{2007ARA&A..Zinnecker, 2012MNRAS.Thompson, 2012ApJ...Kendrew}, associating a radio continuum source, especially \ion{H}{ii} region,  with a molecular cloud catalogue \citep[as in the case of][]{2026MNRAS...Langa} that already has a more robust distance solution offers an alternative path to \ion{H}{ii} region distance assignment. Such “distance by association” can increase the sample of \ion{H}{ii} regions with adopted distances and reduce systematic uncertainty, assuming the association is physically valid.

While previous work has assigned \ion{H}{ii} region distances via different and/or direct kinematic methods \citep[e.g.,][]{1987A&A...Caswell, 2004ApJS..Sewilo, 2006ApJS..Quireza, 2012MNRAS.Urquhart, 2012ApJ...Jones, 2012ApJ...Anderson, 2013A&A...Immer, 2014ApJ...Sato}--for instance, \cite{2012ApJ...Anderson}, from the Green Bank Telescope \ion{H}{ii} Region Discovery Survey (GBT HRDS), resolved distances for 149 HRDS \ion{H}{ii} regions with median uncertainties $\sim$0.5 kpc--our method takes a distinct approach. In this work, we adopt a kinematic‐distance by association methodology: we take \ion{H}{ii} region catalogues (e.g., the WISE catalogue \citep[][]{2014ApJS..Anderson, 2015ApJS..Anderson, 2018ApJS..Anderson}, SMGPS extended source catalogue \citep[][]{2025A&A...Bordiu}) and associate each region with a molecular‐cloud catalogue (specifically the SEDIGISM molecular clouds, which already have kinematic distances assigned \citep[][]{2021MNRAS.Duarte}). Once an \ion{H}{ii} region is associated, we adopt the molecular cloud’s distance as the \ion{H}{ii} region’s distance. With this approach we aim to do the following: First, to increase the number of \ion{H}{ii} regions in which a reliable heliocentric (and hence Galactocentric) distance is assigned. Second, to map the distribution of \ion{H}{ii} regions (via their adopted distances) in the inner Galactic plane, enabling placement in spiral‐arm context. Third, to use the distances and associations to examine environmental relationships of \ion{H}{ii} regions and molecular clouds, in particular the possibility of triggered star formation (via shell morphology, cloud compression, spatial offsets, velocity association). Finally, to assess the reliability, uncertainties and limitations of the association method--including spatial/velocity matching criteria, potential mis‐associations, non-circular motions, and the inherited uncertainties from molecular cloud distances.

We take advantage of the dense molecular-cloud catalogue in the inner Galaxy from SEDIGISM \citep[][]{2021MNRAS.Duarte} and the sensitive, high-resolution and excellent image fidelity of \ion{H}{ii} region sample from SMGPS to systematically associate ionized regions with their natal gas. We believe that this would bypass some of the inherent uncertainties of direct kinematic methods for \ion{H}{ii} regions and allows us to produce a statistically significant, homogeneously-derived dataset specifically designed for environmental studies, such as triggered‐star‐formation investigations \citep[e.g.,][]{1998ASPC..Elmegreen, 2005A&A...Deharveng, 2007MNRAS.Dale, 2009MNRAS.Morgan, 2010A&A...Zavagno}.

In Section \ref{sec: Data} we describe our data and source catalogues--\ion{H}{ii} region catalogues (SMGPS, WISE) and the molecular cloud catalogue (SEDIGISM). In Section \ref{sec: method} we outline our methodology, defining our association criteria (spatial, velocity, angular separation), for assigning distances. In Section \ref{sec: Results} we present results of the distance-assignment: number of \ion{H}{ii} regions with adopted CO-based distances, velocity validation, distance presentation--handling the kinematic distance ambiguity, comparison with previous distance determinations. We also statistically investigate physical properties of associated and unassociated SEDIGISM molecular clouds, and also present moment maps. The relationship between ionising photon flux and distance of the \ion{H}{ii} regions is presented in Section \ref{subsec: photon flux and distance}. In Section \ref{sec: Discussion} we present a general discussion. Finally, we summarise our conclusions in Section \ref{sec: Conclusions}.

\section{Data and source catalogues}
\label{sec: Data}

This analysis relies on three main data surveys and/or catalogues to identify \ion{H}{ii} regions, measure their velocities, and assign kinematic distances by association with molecular clouds.

We use the V2.3 Wide-field Infrared Survey Explorer (WISE) Catalogue\footnote{\url{http://astro.phys.wvu.edu/wise}} of Galactic \ion{H}{ii} Regions \citep{2014ApJS..Anderson, 2015ApJS..Anderson, 2018ApJS..Anderson}, which contains 8416 objects classified as either known (K, 2376 objects), group (G, 635), candidate (C, 1687) or quiet (Q, 3718) type of \ion{H}{ii} regions. The known \ion{H}{ii} regions have detections of radio recombination line (RRL) or H$_\alpha$ emission, which directly trace the ionised gas and provide line-of-sight velocities that may be used with a Galactic rotation curve to infer kinematic distances (subject to near–far ambiguity in the inner Galaxy). On the other hand, the much larger population of candidate and radio quiet types lack RRL measurements-one of the main motivations for our distance-by-association method. The WISE catalogue was created or constructed through both visual and automatic searches of the WISE data that involved the identification of \ion{H}{ii} regions from their mid-infrared (MIR) emission morphology. The data contains WISE 12 $\mu$m and 22 $\mu$m images across the entire Galactic plane within $8^{\circ}$ of the nominal mid-plane, $|b|\leq 8^{\circ}$.

We also use radio continuum data from the SARAO MeerKAT 1.3\,GHz Galactic Plane Survey (SMGPS)\footnote{\url{https://doi.org/10.48479/3wfd-e270}} \citep{2024MNRAS..Goedhart}. The survey observed the Galactic Plane at 1.3\,GHz with an angular resolution of 8 arcseconds and a root-mean-square (RMS) sensitivity of $\sim$10--20 $\mu$Jy/beam. Observations were conducted between July 2018 and March 2020 from the 64-antenna MeerKAT array in the Northern Cape Province of South Africa \citep{Jonas:2018Jr, 2020ApJ...Mauch, 2024MNRAS....Klutse} using the L-band (856--1712 MHz) receiver system with 4096 channels. The survey covers a wide portion of the first, second, and fourth Galactic quadrants ($\ell=2^{\circ}$--$61^{\circ}$, $251^{\circ}$--$358^{\circ}$, $|b|\leq 1.5^{\circ}$). In this work we use the SMGPS moment 0 maps of all tiles within these regions, along with the associated SMGPS catalogue of 16534 extended and diffuse radio sources{\color{red}\footnote{\url{https://doi.org/10.48479/t1ya-na33}}$^{,}$%
\footnote{\url{http://cdsweb.u-strasbg.fr/cgi-bin/qcat?J/A+A/}}$^{,}$%
\footnote{\url{https://cdsarc.cds.unistra.fr/viz-bin/cat/J/A+A/695/A144}}}
\citep{2025A&A...Bordiu}. Of these, $\sim$24 per cent correspond to known Galactic objects, including 3326 \ion{H}{ii} regions, 263 supernova remnants (SNRs), 215 planetary nebulae (PNe), 20 luminous blue variables (LBVs), 7 Wolf-Rayet (WR) stars and 59 objects with multiple associations. The remainder consists of either candidate extragalactic sources ($\sim$33 per cent) or unclassified radio objects ($\sim$43 per cent). Since the SMGPS \ion{H}{ii} region sample in \cite{2025A&A...Bordiu} was originally built directly from a positional cross-match with the WISE catalogue \citep{2014ApJS..Anderson,2015ApJS..Anderson, 2018ApJS..Anderson}, this study focuses exclusively on the SMGPS \ion{H}{ii} region population.

To determine distances through cloud association, we used the molecular catalogue from the SEDIGISM survey \citep{2021MNRAS.Duarte}. The SEDIGISM survey \citep{2017A&A...Schuller, 2021MNRAS.Duarte, 2021MNRAS...Urquhart,2021MNRAS..Schuller} was conducted with the Swedish Heterodyne Facility Instrument \citep[SHFI;][]{2008A&A...Vassilev}-single-pixel instrument of the 12m Atacama Pathfinder Experiment \citep[APEX;][]{2006A&A...Gusten,2009A&A...Schuller,2022A&A...Yang}, between 2013--2017, imaging the $^{13}$CO (2--1) (and C$^{18}$O (2--1), plus other transitions from other molecules) across the Galactic plane between 300 and 17 degrees in longitude, and +/-0.5 degrees in latitude, at $\sim$30-arcsecond resolution. The catalogue used here is the merged catalogue of $^{13}$CO (2--1) molecular clouds as of [Jan.2021]\footnote{\url{https://sedigism.mpifr-bonn.mpg.de/cgi-bin-seg/SEDIGISM_DATABASE.cgi}}, which contains the original clouds from \cite{2021MNRAS.Duarte}, plus additional information and updated distances from \cite{2022A&A...Colombo}, and \citet{2022A&A...Neralwar, 2022A&A...Neralwar_b}. The catalogue was constructed using the Spectral Clustering for Molecular Emission Segmentation ({\sc scimes}) algorithm \citep{2015MNRAS..Colombo, 2019MNRAS...Colombo}, which employs clustering techniques to identify groupings of connected structures in 3D spatial and velocity space, with similar emission properties. 

The SEDIGISM catalogue provides kinematic distances for $\sim$10300 molecular clouds. However, not all of these distances are considered reliable. In the \citet{2021MNRAS.Duarte} catalogue, each SEDIGISM cloud is assigned a reliability flag ($d_\text{reliable}$) that identifies whether the kinematic distance is reliable ($d_\text{reliable}=1$) or unreliable or non-existent ($d_\text{reliable}=0$). Clouds with $d_\text{reliable}=0$ include those with low radial velocities (i.e.~where the systemic velocity is dominated by local motion), unresolved kinematic distance ambiguity, or sources too close in longitude to the Galactic Centre (where the systemic velocity is dominated by proper motion). Our analysis thus proceeds by initially associating \ion{H}{ii} regions with the full SEDIGISM catalogue of 10663 molecular clouds, but restricting the sample to only the subset of 7993 clouds with reliable distances ($d_\text{reliable}=1$) when considering distance dependent quantities.

The datacubes of the emission and cloud masks (each with 0.25\,\kms--wide velocity channels between -200 and 200\,\kms) of $^{13}$CO (2--1) for our analysis were also obtained from the public SEDIGISM web page\footnote{\url{https://sedigism.mpifr-bonn.mpg.de/index.html}}.

\section{Method}
\label{sec: method}

While radio recombination lines (RRLs) provide the most direct and reliable measurements of the line-of-sight velocities of \ion{H}{ii} regions, only a minority of Galactic \ion{H}{ii} regions have detected RRLs because the RRL emission is intrinsically faint and requires relatively strong continuum backgrounds \citep[e.g.,][]{2014ApJS..Anderson, 2015ApJS..Anderson}. Many SMGPS sources therefore lack direct ionised-gas velocity measurements. In such cases, the velocity must be inferred indirectly, most commonly through association with molecular material traced by CO or other dense-gas tracers \citep[e.g.,][]{2009ApJ...Roman-Duval, 2018MNRAS...Urquhart}. To do this, we first took the central approach of assigning systemic velocities to the SMGPS \ion{H}{ii} regions by association with molecular clouds from the SEDIGISM $^{13}$CO (2--1) survey. This is trivial to do for compact embedded radio continuum sources but less so for extended regions with complex morphology. Since molecular clouds provide bright $^{13}$CO (2--1) emission and have robust kinematic distance solutions, they serve as the optimal velocity reference for \ion{H}{ii} regions lacking RRL detections. The whole procedure is detailed in \cite{2026MNRAS...Langa}. To briefly summarise here, the method relies on identifying molecular clouds whose centroid velocities and position–velocity structure are physically consistent with the morphology and projected location of each radio continuum source. In other words, the method identifies the most physically plausible SEDIGISM molecular cloud along each line of sight using spatial overlap, velocity-window selection, and an emission-fraction metric, \fw\,--which is simply the fraction of the total integrated intensity within a velocity window. The velocity windows were determined by first masking the {\sc scimes} assignment cube using the SMGPS \ion{H}{ii} region footprint to isolate associated molecular emission. The velocity windows were then identified as velocity-coherent emission regions consisting of consecutive velocity channels containing non-zero values in the assignment cube, which led to the velocity window's range being defined by the minimum and maximum velocities of its corresponding contiguous channel group. Such windows may contain a single molecular cloud or multiple clouds overlapping in velocity.

The velocity window with highest \fw, within an average $^{13}$CO spectrum (averaged across the entire area of the source), is the best matching window. All the SEDIGISM molecular clouds falling within this best-matching emission window's velocity range and overlapping the SMGPS footprint are considered to be associated with the radio source, hence are referred to as associated clouds. In cases where a single SMGPS \ion{H}{ii} region is associated with multiple SEDIGISM molecular clouds  within the best-matching velocity window, the clouds become part of the associated molecular complex and the SMGPS \ion{H}{ii} region is then assigned the central velocity of the window's range \citep[Table 1,][]{2026MNRAS...Langa}. Thus, for the purpose of deriving physical properties, we treat these as a combined system. For example, all distance-dependent quantities are rescaled to a common distance, defined as the median distance of the associated clouds. Specifically, cloud masses are rescaled before being summed to obtain the total molecular mass of the complex. In some cases for other physical parameters, such as linewidth and virial parameter, we adopt mass-weighted averages of the individual cloud properties. All this is to ensure consistency, making sure that the derived quantities are representative of the dominant molecular component while accounting for the contribution of all associated clouds. In practice, most \ion{H}{ii} regions are found to be associated with a single cloud (see Section \ref{sec: Results}).

We perform the cloud association using the full SEDIGISM molecular cloud catalogue \citep[10663 clouds;][]{2021MNRAS.Duarte}, allowing robust assignment of systemic velocities to the \ion{H}{ii} regions. However, for distance determination, we restrict the adopted kinematic distances to sources where SEDIGISM gave a reliable distance ($d_\text{reliable}=1$). For sources associated with clouds that do not satisfy this criterion, velocities are retained, but distances are excluded from further analysis.
Therefore, like velocities, the mean distance of all the distances of clouds intersecting the best-matching window is assigned to the associated SMGPS \ion{H}{ii} region, hence the adopted distance.

As by definition all of our SMGPS \ion{H}{ii} regions are contained within the WISE \ion{H}{ii} region Catalogue we are thus able to determine radial velocities and distances for WISE \ion{H}{ii} regions. To ensure that our results are not affected by confusion we restrict our final sample to only those objects that have a one-to-one match between SMGPS and WISE. Thus we were able to transfer all the CO-based systemic velocities and kinematic distances derived from the the SMGPS-SEDIGISM association directly to the matched WISE \ion{H}{ii} regions, which we adopt as our final kinematic distances and  systemic velocities to the \ion{H}{ii} regions. We applied the following quality control criteria to obtain the final sample:  (i) unique and reliable object identification, whereby we used only SMGPS \ion{H}{ii} regions with a single, unambiguous object name (since some of the \ion{H}{ii} regions in the SMGPS extended source catalogue have multiple object names); (ii) one-to-one spatial association in which the WISE \ion{H}{ii} region must overlap spatially with the SMGPS extended \ion{H}{ii} region, ensuring that any RRL velocity refers to the same physical object; and (iii) successful molecular-cloud association whereby the SMGPS source must be associated with at least one SEDIGISM molecular cloud through our velocity-window and emission-fraction (\fw) criteria. 

Overall, this procedure yields a sample of $\sim$741 \ion{H}{ii} regions with CO-derived systemic velocities, associating with 684 SEDIGISM molecular clouds. Of these, 640 \ion{H}{ii} regions are associated with molecular clouds that have reliable kinematic distances, corresponding to 572 molecular clouds, and therefore have both CO-based velocities and distances adopted in this work. The remaining 101 radio sources are associated with 112 molecular clouds that do not satisfy the distance reliability criterion; for these sources, CO-derived velocities are retained, but distances are excluded from further analysis.

\section{Results}
\label{sec: Results}

\begin{figure}
    \centering
    \includegraphics[width=\linewidth]{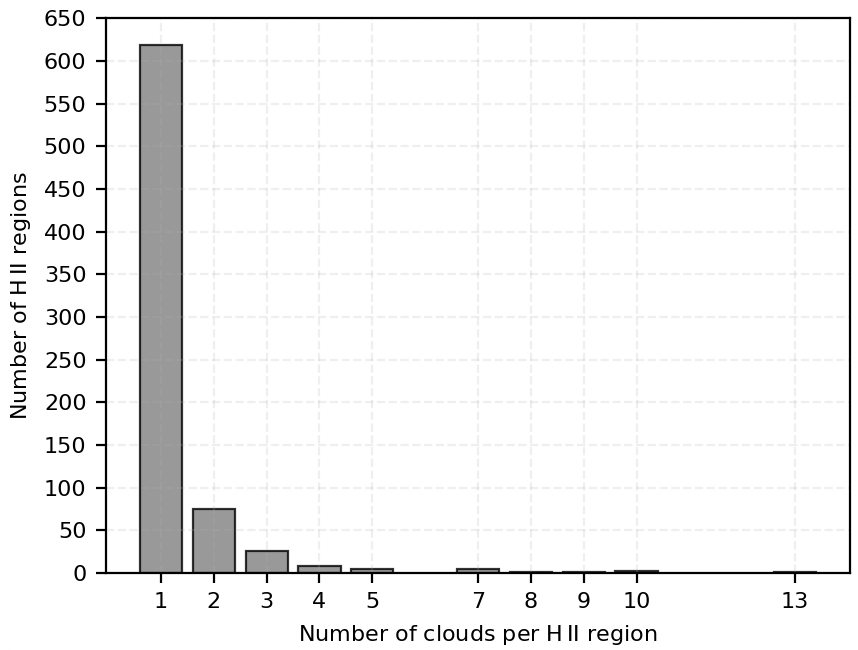}
    \caption{Histogram distribution of the number of SEDIGISM molecular clouds per \ion{H}{ii} region, within the best-matching velocity window, associated with SMGPS extended \ion{H}{ii} regions.}
    \label{fig: histogram_clouds_HII}
\end{figure}

We present our results by first examining the distribution of SEDIGISM molecular cloud associations per SMGPS \ion{H}{ii} region, using the best-matching velocity window which is selected by our emission fraction (\fw) criteria \citep{2026MNRAS...Langa}. This has provided a direct test of how clean and unambiguous the radio-cloud associations are. Figure \ref{fig: histogram_clouds_HII} shows the histogram of cloud-source associations demonstrating that 619 ($\sim$84 per cent) of all \ion{H}{ii} regions have a one-to-one association with a single SEDIGISM molecular cloud with only a minority (16 per cent) containing two or more clouds, with the most extreme cases involving two associations of 2 sources with ten clouds for each, and one source which is linked to thirteen (13) clouds. Thus, the distribution strongly indicates that the SMGPS–SEDIGISM association method is typically unique as most \ion{H}{ii} regions are linked to only one molecular cloud, and the adopted velocity-window and emission fraction criteria are efficient at identifying the correct molecular environment. This high degree of one-to-one matching is a major advantage of the method and significantly increases confidence in the robustness of the assigned velocities.

From the full SMGPS–WISE combined catalogue, our association method identifies 741 \ion{H}{ii} regions with reliable cloud associations. However, as mentioned before, for all analyses involving kinematic distances and derived physical properties, we consider only the subset of 640 \ion{H}{ii} regions associated with SEDIGISM molecular clouds that have reliable kinematic distances. This ensures that the statistical results presented here are not biased by uncertain or poorly constrained distance estimates. Among these radio continuum sources (640), 257 WISE \ion{H}{ii} regions already have published kinematic distances and RRL velocity measurements, while the remaining 483 \ion{H}{ii} regions previously had no distance information available in the literature (although some have RRL velocity measurements) and therefore adopt CO-based distances for the first time (see Table \ref{tab:table1}). Even for the 257 WISE \ion{H}{ii} regions with existing distances, we still compute CO-based kinematic distances. We think that this approach is justified for three main reasons: (i) as demonstrated in Section \ref{subsec: velocity_validation}, it allows for the validation that the SEDIGISM cloud centroid velocities are highly reliable tracers of the radial velocity of the associated \ion{H}{ii} regions; (ii) in terms of beam size, the SEDIGISM survey offers higher angular resolution and better sensitivity than many of the molecular datasets previously used in the literature; and (iii) reassessing these distances ensures a consistent, uniform distance scale across the entire SMGPS–WISE sample. Consequently, the CO-based distances represent a homogeneous and physically motivated revision of the distance estimates for the entire \ion{H}{ii} region sample.

Notably, we also realize that there some \ion{H}{ii} regions that associate with clouds with distance of --1 ($d_\text{flag}=-1$). According to \cite{2021MNRAS.Duarte}, there are 363 molecular clouds presented in the SEDIGISM catalogue with distance --1. These clouds could not be assigned distances since they do not have velocities that fit in the rotation model. We have identified 13 \ion{H}{ii} regions that associate with 13 of such clouds. Such associations have been so far excluded from this study as they will not give us any meaningful interpretation. We also note, however, that a handful of WISE \ion{H}{ii} regions in our sample (4 sources) have distances determined via trigonometric parallax measurements. In this paper, we retain and adopt these parallax-based distances without modification, as they provide the most accurate distance estimates currently available. Importantly, where comparisons are possible, our CO-based distance assignments are consistent with the parallax measurements, reinforcing the reliability of the association method.

\subsection{Comparison of RRL and CO velocities}
\label{subsec: velocity_validation}

Because inner-Galaxy lines of sight contain multiple molecular components with overlapping velocity ranges and non-trivial spatial substructure \citep{2014A&A...Hou, 2022A&A...Colombo}, it is essential to validate that our CO-based velocity assignments correctly recover the true radial velocity of the \ion{H}{ii} regions.

As mentioned in Section \ref{sec: Data}, all the SMGPS extended \ion{H}{ii} regions have counterparts in the WISE Catalogue of Galactic \ion{H}{ii} Regions \citep[][]{2014ApJS..Anderson, 2015ApJS..Anderson, 2018ApJS..Anderson}. However, only a subset of these WISE \ion{H}{ii} regions has published independent systemic velocities measured directly from radio recombination lines (RRLs). RRLs trace the ionized gas, whereas CO emission traces the surrounding molecular environment; despite this physical distinction, the molecular and ionized gas components in genuine \ion{H}{ii} region–cloud systems are expected to share similar systemic velocities. Therefore, these WISE RRL velocity measurements, which represent a direct measurement of the source’s systemic motion, provide an ideal external benchmark. Out of the 741 WISE \ion{H}{ii} regions, only 369 have RRL velocity measurements, and to confirm the velocity validation, we plotted the SMGPS CO-derived velocity versus the reliable WISE RRL velocity where we used only the 329 WISE \ion{H}{ii} regions with a single, unambiguous RRL velocity measurement. This sample is an order of magnitude larger than the $\approx$20-object verification performed in our earlier SMGPS pilot study--G342.5-tile analysis \citep{2026MNRAS...Langa}, enabling a Galaxy-wide test of our velocity-assignment methodology. 

\begin{figure}
    \centering
    \includegraphics[width=\linewidth]{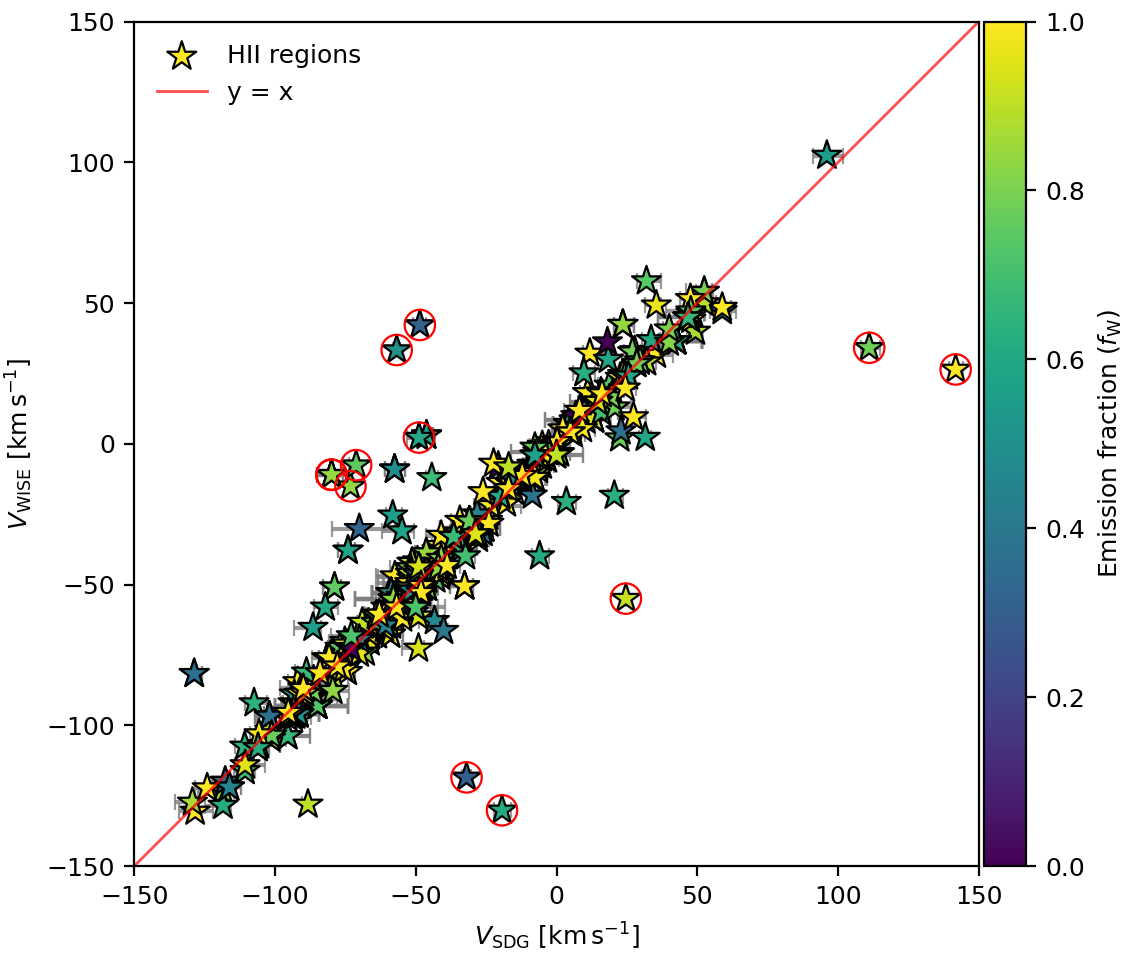}
    \caption{Scatter plot of the CO-derived centroid velocities of matched \ion{H}{ii} regions from the SMGPS extended sources versus the systemic velocities of \ion{H}{ii} regions from the WISE catalogue of Galactic \ion{H}{ii} regions. The colour bar shows the emission fraction \fw\ of the associations between the interacting SMGPS extended \ion{H}{ii} regions with the associated SEDIGISM molecular clouds from the best spectral emission windows. The red line indicates the one-to-one relation. The highlighted data points with red circles are sources with absolute velocity discrepancy > 50 \kms.}
    \label{fig:SMGPS_WISE}
\end{figure}

Figure \ref{fig:SMGPS_WISE} compares the CO-derived velocities V$_\text{SDG}$ with the WISE RRL velocities V$_\text{WISE}$, showing the data points that tightly cluster around the one-to-one line across the entire velocity range. This indicates a strong correlation and correspondence between the two velocity measurements, and hence the molecular gas and ionized gas systemic motions. The mean absolute velocity difference is 7.67$\pm$14.85 \kms\,, showing that the broad standard deviation is a reflection of a small number of large outliers. Only a minority of \ion{H}{ii} regions show discrepancies beyond 30 \kms\, and 10 of them show extreme velocity outlier> 50 \kms\, (highlighted using red circles in Figure \ref{fig:SMGPS_WISE}). Moreover, \ion{H}{ii} regions with high emission fraction \fw\, values are found closest to the one-to-one relation, a confirmation that the emission fraction metric is an effective indicator of robust velocity assignment. Overall, the large sample shows good velocity agreement, with a median absolute velocity difference of only 3.46 \kms\, between the CO-- and RRL--based measurements.

\begin{landscape}
\begin{figure}
    \centering
    \includegraphics[width=\linewidth]
    {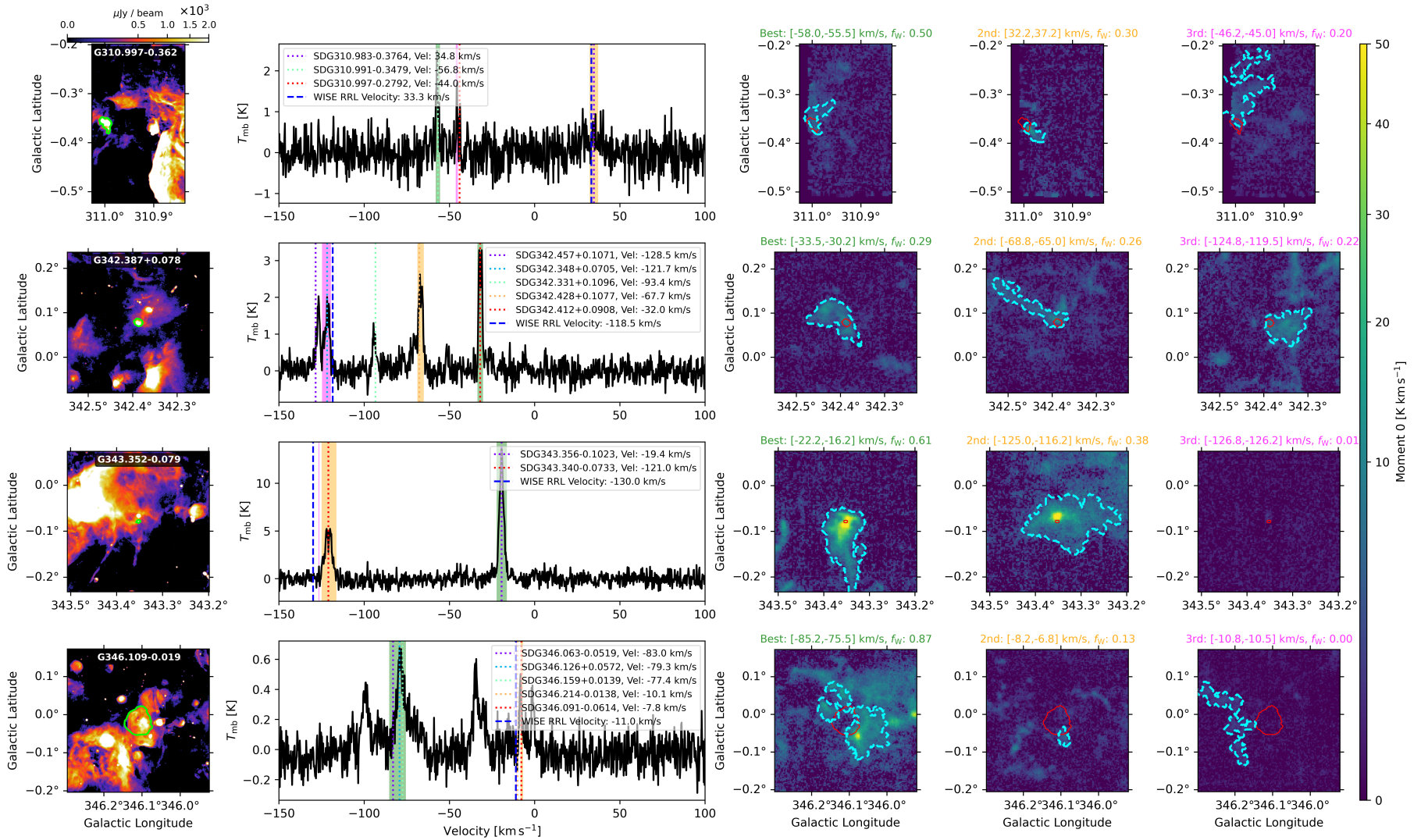}
    \caption{An illustration of the SMGPS-SEDIGISM velocity assignment approach for four \ion{H}{ii} regions. For each row (source), the left panel shows the SMGPS 1.3 GHz intensity image with the \ion{H}{ii} region boundary \citep{2025A&A...Bordiu} outlined in lime tracing its emission. The middle panel shows the mean SEDIGISM $^{13}$CO (2--1) spectrum extracted over the same region, with the three most significant velocity-coherent windows highlighted--best, second, and third ranked by emission fraction \fw\,--in green, orange and magenta, respectively. The spectra show the spatially averaged $^{13}$CO emission extracted only within the \ion{H}{ii} region contour (lime) shown in the left panels. The velocities of individual SEDIGISM molecular clouds, intersecting each window, are marked in vertical dashed lines. Dashed blue vertical lines indicate the radio recombination line (RRL) velocities \citep[WISE catalogue;][]{2014ApJS..Anderson, 2015ApJS..Anderson, 2018ApJS..Anderson}. The right panels display the corresponding moment-0 maps for the top three windows, with the \ion{H}{ii} region outline \citep{2025A&A...Bordiu} overlaid and shown in red polygon, while the masks of the associated SEDIGISM molecular clouds are shown as dashed cyan contours. However, some low-significance velocity windows contain sparse or fragmented {\sc scimes}-assigned voxels, leading to weak or absent cloud contours in the corresponding moment maps. The moment maps are integrated over the corresponding velocity windows across the full spatial field, thereby revealing the larger-scale molecular cloud morphology associated with each velocity component. For each window, the velocity range and emission fraction \fw\, are indicated. These examples demonstrate how the procedure identifies the most physically plausible cloud-radio (\ion{H}{ii} region) cross-match even in complex and multi-component lines of sight.}
    \label{fig: HII_regions_plots}
\end{figure}
\end{landscape}

We examined all clouds with large velocity discrepancies ($\ge$ 30 km\,s$^{-1}$) to investigate why there is a disagreement between the RRL and CO velocity. We found that the discrepancies largely arise when the \ion{H}{ii} regions are associated with multiple SEDIGISM clouds with similar emission fractions. Examples of these are shown in Figure \ref{fig: HII_regions_plots}. In all cases that we examined we found that there is CO emission close to the RRL velocity, although it is not the component with the highest emission fraction and does not often agree morphologically with the radio emission. This can be seen for the top two \ion{H}{ii} regions in Figure \ref{fig: HII_regions_plots}.

The remaining \ion{H}{ii} regions in the bottom two rows of Figure \ref{fig: HII_regions_plots} show cases where the RRL velocities do not agree with the strongest CO emission. \ion{H}{ii} region G343.352$-$0.079 (in the 3rd row) presents a spectrum with two strong emission fractions (0.61 and 0.38). These are likely to be two line-of-sight molecular clouds, each with emission peaks close to the \ion{H}{ii} region. The strongest component has the highest emission fraction, although it is difficult to distinguish them morphologically. However, \ion{H}{ii} region G346.109$-$0.019 (in the 4th row) clearly shows that the morphologically best association (with an emission fraction of 0.87) does not agree with the RRL velocity. The RRL velocity measured by \citet{1987A&A...Caswell} corresponds to weak CO emission and thus the RRL emission is likely to be from the diffuse ionised medium.

\begin{landscape}
\begin{figure}
    \vspace*{2.5cm}
    \centering
    \includegraphics[width=\linewidth]{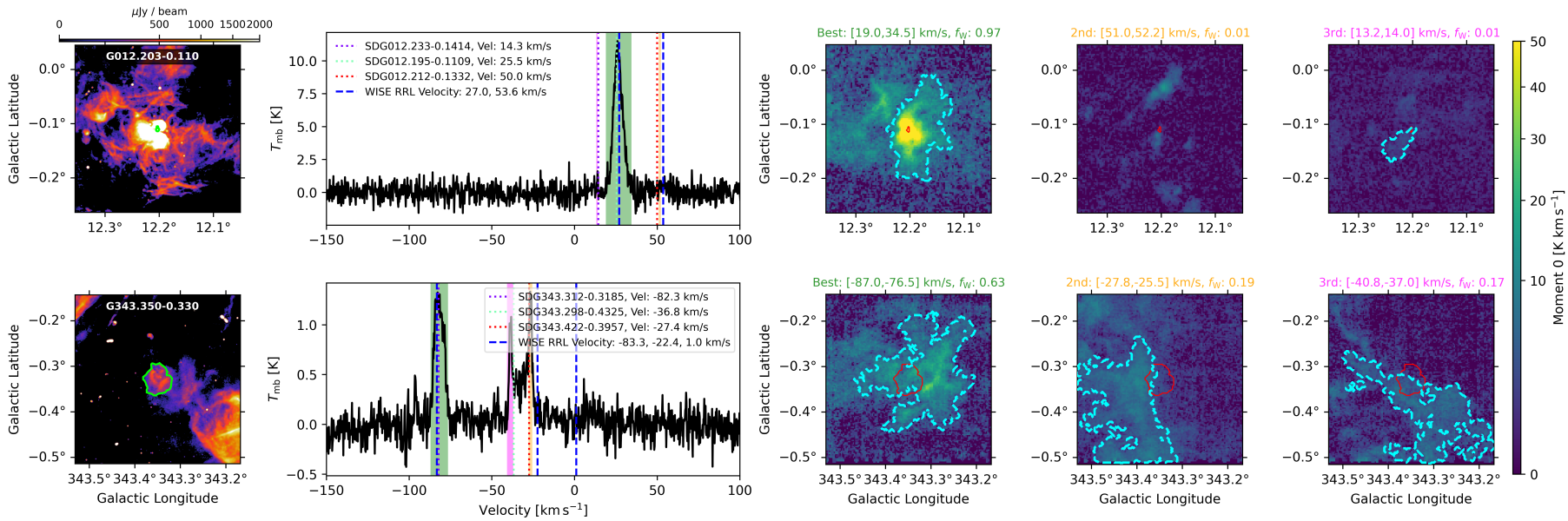}
    \caption{Same as Figure \ref{fig: HII_regions_plots} for the \ion{H}{ii} regions G012.203-0.110 and G343.350-0.330.}
    \label{fig: HII_with_2_n_3vels_plots}
\end{figure}
\end{landscape}

Some of the RRL-CO velocity discrepancies occur for \ion{H}{ii} regions associated with multiple RRL velocity components. Here our emission fraction method of associating SEDIGISM clouds shows its strength in using CO and radio continuum morphology to select the correct velocity component for the \ion{H}{ii} region. We show two examples of this in Figure \ref{fig: HII_with_2_n_3vels_plots}. The first example is  G012.203$-$0.110 was assigned two RRL velocities of 27.0 and 53.6 \kms\ measured by \cite{1989ApJS...Lockman}. The strongest emission fraction component is clearly morphologically associated with the \ion{H}{ii} region and thus we can confidently assign a velocity of 25.5 \kms\ to this region. Similarly,  our second example of \ion{H}{ii} region G343.350$-$0.330  has three RRL velocity components listed in \cite{2018ApJS..Anderson}. We can show that the component most clearly associated with the morphology of ionised and molecular gas is that at $-$82.3 \kms.

In our further analysis we have chosen to retain the CO velocities for all \ion{H}{ii} regions, regardless of any discrepancy between their RRL and CO-derived velocity. Our reasoning behind this is two-fold. Firstly, only half of our total sample of 741 \ion{H}{ii} regions have measured RRL velocities. Using CO-derived velocities for the entire sample thus  allows statistical comparisons that are free of selection effects caused by using different velocity derivations. Secondly, our emission fraction method represents a major improvement in previous velocity-matching methods which have typically used single position spectra \citep[often with beams approaching an arcminute or greater, e.g.][]{1987A&A...Caswell, 2001ApJ...Dame} and matched the brightest line to the continuum detection. Our emission fraction method brings in morphology to the matching process for the first time and we are confident that our method improves the security of the cloud-\ion{H}{ii} associations. Furthermore, Figures \ref{fig: histogram_clouds_HII} and \ref{fig:SMGPS_WISE} show that the vast majority of \ion{H}{ii} regions are associated with only a single SEDIGISM cloud and that the vast majority of RRL velocities agree with the SEDIGISM derived velocity. Any discrepancies in our ample are likely to only affect a small number of \ion{H}{ii} regions.

As we have observed from Figure \ref{fig: HII_with_2_n_3vels_plots}, it is clear that all WISE \ion{H}{ii} regions with multiple RRL measurements should adopt the CO-based velocity derived from our SEDIGISM association within the best-matching velocity window. This provides a unique, physically justified systemic velocity for each \ion{H}{ii} region and therefore removes their kinematic ambiguity entirely, particularly in crowded sight lines, thereby improving the accuracy and reliability of Galactic \ion{H}{ii} region distance determinations to effectively recover objects that would otherwise remain excluded from Galactic structure analyses.

We discuss the effects upon the derived distances  in Section \ref{subsec: dist_comparison}.

\begin{table*}
\centering
\caption{Cross-match of the SMGPS extended radio continuum sources, the SEDIGISM molecular clouds and the WISE \ion{H}{ii} regions. Column 1 presents the SMGPS extended \ion{H}{ii} region name in IAU format. Columns 2--3 present cloud name, and systemic velocity for the associated SEDIGISM molecular cloud (reference for all the velocities for the associated SEDIGISM molecular clouds (SDG $\mathrm{v_{lsr}}$ reference) is \protect\cite{2021MNRAS.Duarte}). Column 4 gives the cloud's velocity range for the best-matching velocity window where all the associated clouds intersect, column 5 presents the mean centroid velocity of all cloud's velocities within the best-matching velocity window. Columns 6--8 present kinematic distance assigned to the SEDIGISM molecular cloud, mean distance for distances of the clouds falling within the best-matching velocity window, and the reference for the distance of SEDIGISM molecular cloud. Column 9 presents an emission fraction (\fw) for the best matching velocity window. Columns 10--12 give name, RRL velocity measurement, and the  reference for the RRL velocity for the WISE \ion{H}{ii} region. Column 13 gives CO-derived systemic velocity assigned to the \ion{H}{ii} region as a result of the cloud association. Columns 14--16 give distance measurement, distance method, and the reference for the distance for the WISE \ion{H}{ii} region. Columns 17--18 give the CO-based adopted distance to the \ion{H}{ii} regions, and the cloud association method used to assign distances.}
\resizebox{\textwidth}{!}{
\begin{tabular}{cccccccccc}
\hline
\hline
SMGPS & SEDIGISM & cloud v$_\text{lsr}$ & v$_\text{lsr}$ range & Mean SDG v$_\text{lsr}$ & cloud dist & Mean dist & Dist$_\text{SDG}$ reference & Emission fraction\\
(iauName) & (Cloud Name) & (\kms) & (\kms) & (\kms) & (kpc) & (kpc) & - & (\fw)\\
(1) & (2) & (3) & (4) & (5) & (6) & (7) & (8) & (9) \\
\hline
G001.980+00.188 & SDG001.989+0.1942 & 131.29 & 126.25, 138.50 & 131.29 & 8.33 & 8.33 ± 1.12 & {\citealt{2021MNRAS.Duarte}} & 0.9905 \\
G009.182+00.026 & SDG009.211+0.0573 & 31.52 & 29.25, 33.50 & 31.52 & 3.97 & 3.97 ± 0.70 & {\citealt{2021MNRAS.Duarte}} & 0.5883 \\
G012.775+00.332 & SDG012.801+0.4129 & 18.71 & 15.25, 21.00 & 18.71 & 2.16 & 2.16 ± 0.81 & {\citealt{2021MNRAS.Duarte}} & 0.9721 \\
G012.908-00.279 & SDG012.840-0.2041 & 35.48 & 31.00, 41.25 & 35.48 & 2.63 & 2.63 ± 0.26 & {\citealt{2021MNRAS.Duarte}} & 1.0000 \\
G013.384+00.065 & SDG013.380+0.0649 & 15.76 & 10.75, 20.00 & 15.76 & 1.82 & 1.82 ± 0.82 & {\citealt{2021MNRAS.Duarte}} & 1.0000 \\
G013.996-00.203 & SDG013.960-0.1733 & 40.03 & 37.75, 43.25 & 40.03 & 3.69 & 3.69 ± 0.54 & {\citealt{2021MNRAS.Duarte}} & 0.8016 \\
G016.767+00.022 & SDG016.761+0.0178 & 34.35 & 28.25, 37.25 & 34.35 & 2.97 & 2.97 ± 0.56 & {\citealt{2021MNRAS.Duarte}} & 0.9270 \\
G304.288-00.300 & SDG304.298-0.3308 & 33.33 & 30.75, 33.50 & 33.33 & 11.9 & 11.9 ± 0.67 & {\citealt{2021MNRAS.Duarte}} & 1.0000 \\
G311.478-00.391 & SDG311.471-0.3935 & -50.99 & -53.00, -48.25 & -50.99 & 3.92 & 3.92 ± 0.93 & {\citealt{2021MNRAS.Duarte}} & 0.8322 \\
G319.164-00.422 & SDG319.159-0.4138 & -23.22 & -29.25, -12.50 & -23.22 & 11.15 & 11.15 ± 0.51 & {\citealt{2021MNRAS.Duarte}} & 1.0000 \\
G320.339-00.177 & SDG320.324-0.1895 & -8.32 & -13.00, -3.25 & -8.32 & 0.55 & 0.55 ± 0.53 & {\citealt{2022A&A...Colombo}} & 1.0000 \\
G328.168-00.203 & SDG328.183-0.1994 & -97.78 & -98.75, -95.25 & -97.78 & 7.09 & 7.09 ± 1.01 & {\citealt{2021MNRAS.Duarte}} & 1.0000 \\
G328.341-00.481 & SDG328.185-0.4643 & -44.51 & -48.00, -39.25 & -44.51 & 2.76 & 2.76 ± 0.46 & {\citealt{2021MNRAS.Duarte}} & 1.0000 \\
G329.654-00.484 & SDG329.632-0.4529 & -34.48 & -36.75, -34.50 & -34.48 & 2.23 & 2.23 ± 0.48 & {\citealt{2021MNRAS.Duarte}} & 1.0000 \\
G333.128-00.436 & SDG333.312-0.3480 & -50.93 & -63.50, -47.25 & -50.93 & 3.24 & 3.24 ± 0.44 & {\citealt{2021MNRAS.Duarte}} & 0.9884 \\
G333.285-00.395 & SDG333.312-0.3480 & -50.93 & -58.75, -47.00 & -50.93 & 3.24 & 3.24 ± 0.44 & {\citealt{2021MNRAS.Duarte}} & 0.9809 \\
G333.676-00.436 & SDG333.663-0.4440 & -2.83 & -7.75, 0.00 & -2.83 & 14.77 & 14.77 ± 0.68 & {\citealt{2021MNRAS.Duarte}} & 1.0000 \\
G340.050-00.232 & SDG340.024-0.1786 & -52.17 & -61.75, -42.50 & -52.17 & 3.72 & 3.72 ± 0.45 & {\citealt{2021MNRAS.Duarte}} & 0.9977 \\
G340.248-00.371 & SDG340.301-0.3954 & -49.04 & -55.00, -47.25 & -49.04 & 3.57 & 3.57 ± 0.46 & {\citealt{2021MNRAS.Duarte}} & 0.9413 \\
G342.355-00.052 & SDG342.347+0.0118 & -6.20 & -8.25, -3.00 & -6.20 & 0.64 & 0.64 ± 0.77 & {\citealt{2022A&A...Colombo}} & 0.3586 \\
G342.395+00.200 & SDG342.395+0.2009 & -18.40 & -19.75, -16.50 & -18.40 & 14.18 & 14.18 ± 0.68 & {\citealt{2021MNRAS.Duarte}} & 0.8809 \\
G343.721-00.222 & SDG343.756-0.1620 & -26.11 & -31.00, -23.50 & -26.11 & 2.44 & 2.44 ± 0.63 & {\citealt{2021MNRAS.Duarte}} & 1.0000 \\
G346.544+00.061 & SDG346.532+0.0945 & 2.21 & 3.00, 4.25 & 2.21 & 16.77 & 16.77 ± 1.20 & {\citealt{2021MNRAS.Duarte}} & 0.3578 \\
G352.395-00.065 & SDG352.396-0.0794 & -89.93 & -94.25, -86.00 & -89.93 & 7.80 & 7.80 ± 0.36 & {\citealt{2021MNRAS.Duarte}} & 1.0000 \\
G353.279-00.476 & SDG353.265-0.4750 & -9.63 & -12.00, -7.25 & -9.63 & 1.94 & 1.94 ± 1.30 & {\citealt{2021MNRAS.Duarte}} & 1.0000 \\

$\vdots$ & $\vdots$ & $\vdots$ & $\vdots$ & $\vdots$ & $\vdots$ & $\vdots$ & $\vdots$ & $\vdots$\\
\hline\hline
\end{tabular}
}

\vspace{0.3cm}
\centering
\caption*{continued}
\resizebox{\textwidth}{!}{
\begin{tabular}{ccccccccc}
\hline
\hline
WISE Name & WISE$_\text{VLSR}$ 
& WISE$_\text{VLSR}$ reference & Adopted SMGPS$_\text{SDG}$ vel & WISE Dist & Dist$_\text{WISE}$. Method 
& Dist$_\text{WISE}$. reference & Adopted Dist & Dist$_\text{SDG}$. Method\\
- & (\kms) & - & (\kms) & (kpc) & - & - & (kpc) & -\\
(10) & (11) & (12) & (13) & (14) & (15) & (16) & (17) & (18) \\
\hline
G001.975+00.172 & - & - & 131.29 & - & - & - & $-$ $-$ $-$ & - \\
G009.178+00.043 & 2.2 & {\citealt{2011ApJS...Anderson}} & 31.52 & - & - & - & 3.97 ± 0.70 & cloud association \\
G012.774+00.330 & 15.8 & {\citealt{2004ApJS..Sewilo}} & 18.71 & 13.96 ± 0.33 & H2CO/Kin & {\citealt{2004ApJS..Sewilo}} & 2.16 ± 0.81 & cloud association \\
G012.907-00.277 & 31.7 & {\citealt{2004ApJS..Sewilo}} & 35.48 & 2.45 ± 0.15 & Parallax &{\citealt{2013A&A...Immer}} & 2.63 ± 0.26 & cloud association \\
G013.384+00.066 & 18.3 & {\citealt{1989ApJS...Lockman}} & 15.76 & - & - & - & 1.82 ± 0.82 & cloud association \\
G014.000-00.206 & 40.7 & {\citealt{2015ApJ...Anderson_b}} & 40.03 & - & - & - & 3.69 ± 0.54 & cloud association \\
G016.770+00.019 & - & - & 34.35 & - & - & - & 2.97 ± 0.56 & cloud association \\
G304.288-00.300 & - & - & 33.33 & - & - & - & 11.90 ± 0.67 & cloud association \\
G311.484-00.396 & - & - & -50.99 & - & - & - & 3.92 ± 0.93 & cloud association \\
G319.164-00.421 & -22 & {\citealt{1987A&A...Caswell}} & -23.22 & 11.06 ± 0.37 & H2CO/Kin & {\citealt{1987A&A...Caswell}} & 11.15 ± 0.51 & cloud association \\
G320.343-00.180 & -11 & {\citealt{1987A&A...Caswell}} & -8.32 & 12.00 ± 0.34 & H2CO/Kin & {\citealt{1987A&A...Caswell}} & 0.55 ± 0.53 & cloud association \\
G328.168-00.203 & - & - & -97.78 & - & - & - & 7.09 ± 1.01 & cloud association \\
G328.355-00.485 & - & - & -44.51 & - & - & - & 2.76 ± 0.46 & cloud association \\
G329.656-00.486 & - & - & -34.48 & - & - & - & 2.23 ± 0.48 & cloud association \\
G333.129-00.439 & -47.2 & {\citealt{2021ApJS..Wenger}} & -50.93 & 2.95 ± 0.25 & HIEA/Kin & {\citealt{2012ApJ...Jones}} & 3.24 ± 0.44 & cloud association \\
G333.288-00.399 & -52.1 & {\citealt{2021ApJS..Wenger}} & -50.93 & 3.15 ± 0.22 & HIEA/Kin & {\citealt{2012MNRAS.Urquhart}} & 3.24 ± 0.44 & cloud association \\
G333.681-00.441 & -0.5 & {\citealt{2021ApJS..Wenger}} & -2.83 & 14.59 ± 0.29 & HIEA/Kin & {\citealt{2012ApJ...Jones}} & 14.77 ± 0.68 & cloud association \\
G340.051-00.231 & -50.4 & {\citealt{2021ApJS..Wenger}} & -52.17 & 3.54 ± 0.22 & HIEA/Kin & {\citealt{2012ApJ...Jones}} & 3.72 ± 0.45 & cloud association \\
G340.247-00.373 & -72.7 & {\citealt{2015ApJ...Anderson_b}} & -49.04 & 3.90 ± 0.24 & HIEA/Kin & {\citealt{2012MNRAS.Urquhart}} & 3.57 ± 0.46 & cloud association \\
G342.354-00.047 & -4.4 & {\citealt{2021ApJS..Wenger}} & -6.20 & 15.24 ± 0.32 & HIEA/Kin & {\citealt{2012ApJ...Jones}} & 0.64 ± 0.77 & cloud association \\
G342.395+00.199 & - & - & -18.40 & - & - & - & 14.18 ± 0.68 & cloud association \\
G343.721-00.223 & - & - & -26.11 & - & - & - & 2.44 ± 0.63 & cloud association \\
G346.546+00.061 & - & - & 2.21 & - & - & - & 16.77 ± 1.20 & cloud association \\
G352.393-00.066 & -87 & {\citealt{1989ApJS...Lockman}} & -89.93 & - & - & - & 7.80 ± 0.36 & cloud association \\
G353.277-00.480 & - & - & -9.63 & - & - & - & $-$ $-$ $-$ & - \\

$\vdots$ & $\vdots$ & $\vdots$ & $\vdots$ & $\vdots$ & $\vdots$ & $\vdots$ & $\vdots$ & $\vdots$\\
\hline\hline
\end{tabular}
}
\begin{minipage}{18cm}
\vspace{0.2cm}
\footnotesize \textbf{Notes:} Sources whose adopted distances (under the column Adopted Dist) are marked '$-$ $-$ $-$' are those whose counterpart SEDIGISM molecular clouds have unreliable distance flag ( $d_\text{reliable}=0$), hence the distances are not assigned and so does the distance method. A full electronic version of this table has been added as supplementary material. It is in the same format as Table \ref{tab: SMGPS_SED_WISE_merged_catalogue}.
\end{minipage}
\label{tab:table1}
\end{table*}

\subsection{Distance comparison}
\label{subsec: dist_comparison}

Here, we demonstrate the accuracy, consistency, and reliability of our adopted CO-based distances. Therefore, to achieve that, we compare the distances with the already published WISE kinematic distances for the 257 \ion{H}{ii} regions that have both measurements available.

\begin{figure}
    \centering
    \includegraphics[width=\linewidth]{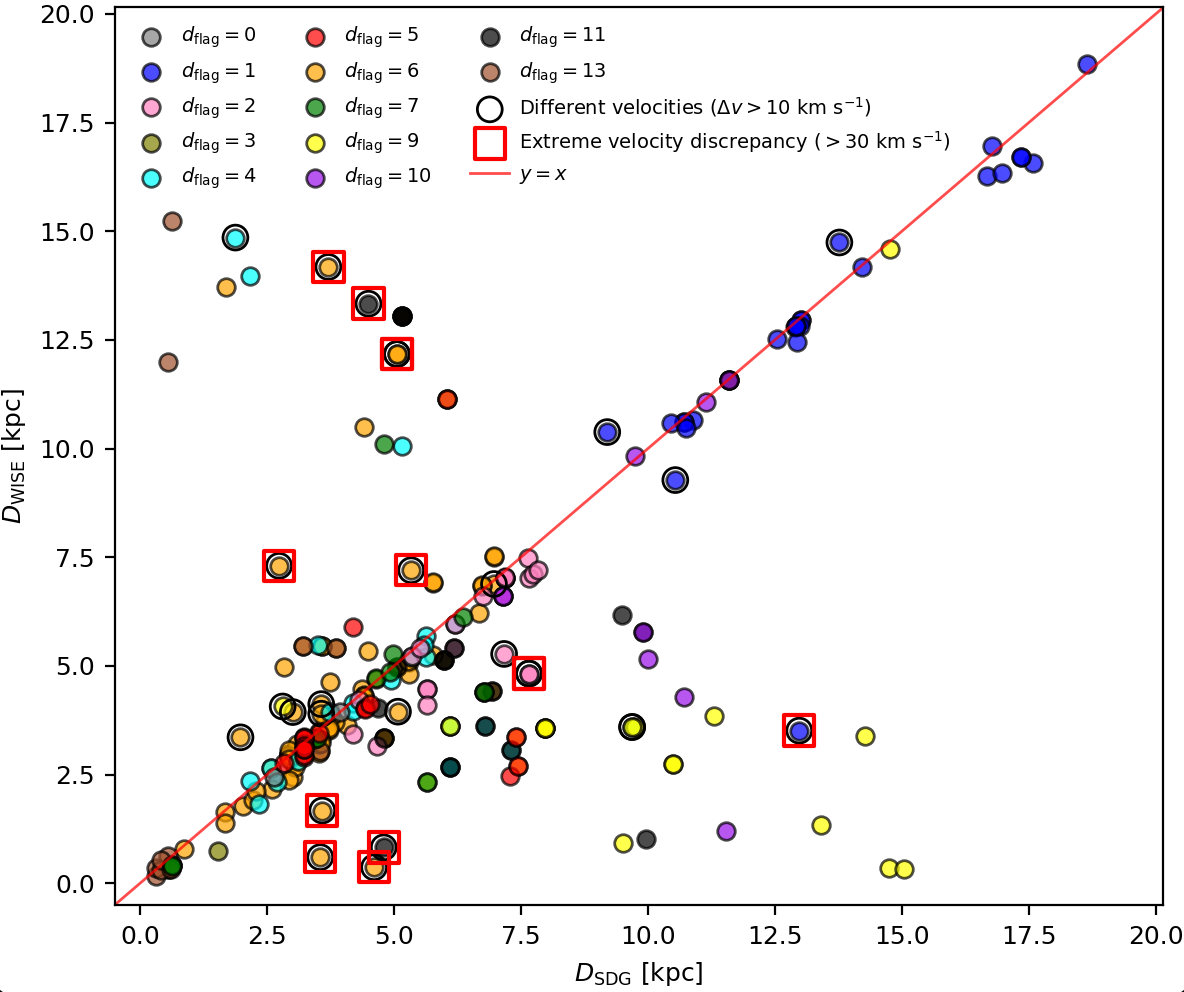}
    \caption{Scatter plot of the adopted CO-based distances versus WISE kinematic distances for 257 \ion{H}{ii} regions with reliable distance determination ($d_{\mathrm{reliable}} = 1$). Data points are colour-coded according to the SEDIGISM distance flag \citep[$d_{\mathrm{flag}}$,][]{2021MNRAS.Duarte}, which describes the method used to resolve the kinematic distance ambiguity: $d_{\mathrm{flag}}=0$--exact maser parallax distance, $d_{\mathrm{flag}}=1$--no distance ambiguity, $d_{\mathrm{flag}}=2$--tangent distance, $d_{\mathrm{flag}}=3$--dark cloud (near distance), $d_{\mathrm{flag}}=4$--IRDC (near distance), $d_{\mathrm{flag}}=5$--literature HISA (near distance), $d_{\mathrm{flag}}=6$--direct HISA measurement (near distance), $d_{\mathrm{flag}}=7$--ATLASGAL source at near distance, $d_{\mathrm{flag}}=9$--size-linewidth scatter \citep[near or far distance,][]{1981MNRAS...Larson, 1987ApJ...Solomon}, $d_{\mathrm{flag}}=10$--ATLASGAL source at far distance, $d_{\mathrm{flag}}=11$--extinction (near or far distance), and $d_{\mathrm{flag}}=13$--spiral arm distance. Lower $d_{\mathrm{flag}}$ values generally correspond to more robust distance determinations, while higher values indicate increasing uncertainty. Sources with discrepancies between CO-derived and WISE RRL velocities are highlighted: open circles mark sources with $\Delta v > 10\,$\kms, while red squares indicate extreme discrepancies with $\Delta v > 30\,$\kms. The red line indicates the one-to-one relation.}
    \label{fig: distance_comparison}
\end{figure}

Figure \ref{fig: distance_comparison} shows the comparison between the distances of the WISE \ion{H}{ii} regions and our adopted CO-based distances for the radio continuum sources. We notice that a large fraction of the data points fall close to the one-to-one relation, hence demonstrating broad consistency between the two assignments and confirming that our molecular cloud associations typically trace the true radial velocity of the radio sources--yielding consistent kinematic distances. However, a subset of sources deviates significantly from the equality line, showing a huge difference in distance measurements. One of the reasons linked to this is due to the velocity discrepancies (also see Figure \ref{fig:SMGPS_WISE}) between the CO-based velocities and the WISE RRL velocities. Some \ion{H}{ii} regions exhibit large velocity discrepancies ($>30-50$\,\kms) and therefore this propagates directly to distance differences.  In that effect, Figure \ref{fig: distance_comparison} presents 28 \ion{H}{ii} regions (marked with open circles) that exhibit differences greater than 10\,\kms, of which 11 sources (indicated with red squares) show extreme velocity discrepancy exceeding 30--50\,\kms. Approximately twenty ($\sim 20$) of these sources, including all the 11 (velocity > 30\,\kms), are clearly associated with some of the largest deviations from the equality line, confirming that velocity mismatches directly propagate into distance discrepancies. Nevertheless, as observed, the number of sources with significant differences in kinematic distance is still larger than the number of extreme velocity discrepancies, indicating that velocity differences alone are not sufficient to explain all outliers.

A further major contributor to the distance offsets is the kinematic distance ambiguity (KDA) that affects all inner Galaxy sources where the rotation curve method gives two possible heliocentric near and far distances \citep{2022A&A...Colombo}. From the SEDIGISM catalogue, kinematic distances of the clouds are flagged ($d_{\mathrm{flag}}$) to give the method description (see caption of Figure \ref{fig: distance_comparison}) which was used to resolve the kinematic distance ambiguity \citep{2021MNRAS.Duarte}. In general, higher $d_{\mathrm{flag}}$ values indicate increasing uncertainty in resolving KDA, while lower values correspond to more robust distance determinations. Therefore, in Figure \ref{fig: distance_comparison}, we find that several of the largest distance deviations ($\sim 24$ sources) are associated with higher $d_{\mathrm{flag}}$ values (e.g. $d_{\mathrm{flag}}\geq9$), which are related to less reliable KDA solution methods such as size-linewidth relations \citep[e.g.][]{1981MNRAS...Larson, 1987ApJ...Solomon} or extinction-based estimates \citep[e.g.][]{2006A&A...Marshall}, which are intrinsically less reliable than direct or geometrically constrained methods. This further supports the interpretation that the largest discrepancies arise from intrinsic uncertainties in the distance assignment methods that eventually contribute to the observed scatter. We also note that none of our distances for the comparison have $d_{\mathrm{flag}}=8$ which represents Solomon distance to Galactic plane (GP) \citep[near distance,][]{1987ApJ...Solomon}, and $d_{\mathrm{flag}}=12$ representing cases with distance ambiguity not resolved (defaulted to far). 

The WISE catalogue is not an exception for adopting such distance measurements as many distances were assigned using rotation curve assumptions. In Figure \ref{fig: distance_comparison}, a substantial fraction of the remaining outliers ($\sim 16$ sources) follows a characteristic distribution that reflects near-far flips or swaps as there is a span from top-left through bottom right of the figure, a feature consistent with near-far distance inversions in the inner Galaxy. The final reason is that the presence of complex molecular environments where multiple clouds intersect the same velocity window, multi-component CO spectra, or unrelated foreground/background molecular structures can bias distances, particularly in the inner Galaxy.

Apart from these large distance outliers, we note that there are small changes in the distance itself, perhaps as a result of systemic shifts. This may arise from differences in the Galactic rotation curve models used to compute the kinematic distances for the WISE sources and SEDIGISM clouds. Most of the WISE kinematic distances were computed using earlier rotation curve models, such as those of the \cite{1986PhDT.........9Brand}, \cite{1993A&A...Brand} and \cite{2009ApJ...Anderson} \citep[e.g.][]{2014ApJS..Anderson, 2015ApJS..Anderson, 2018ApJS..Anderson, 2018MNRAS...Urquhart}. In contrast, the SEDIGISM kinematic distances were derived using the \cite{2016ApJ...Reid} Galactic rotation curve model, which is the more recent curve, constructed using maser parallax distance measurements. The model also adopts the revised Galactic rotation curve parameters such as the distance from the Sun to the Galactic Centre, $R_0 = 8.34$ kpc and a Solar
circular rotation speed, $V_0 = 240$\,\kms. Therefore, the differences between these Galactic rotation curves are expected to introduce small systemic shifts in the derived distances, but are unlikely to produce the large distance discrepancies observed in Figure \ref{fig: distance_comparison}. Additional factors, such as the use of lower-resolution molecular data \citep{2001ApJ...Dame}, or telescopes with large beams that blended unrelated molecular components within single beam averaged RRL spectrum \citep[e.g.,][]{1987A&A...Caswell, 1989ApJS...Lockman}, can further contribute to uncertainties in the WISE distance estimates.

Overall, despite the fact that there are concerns and deviations raised, more than half of the matched \ion{H}{ii} regions still demonstrate strong correlation between CO-based and WISE distances. The remaining discrepancies can be attributed to a combination of velocity mismatches, differing methods for resolving the kinematic distance ambiguity, and the kinematic distance ambiguity affecting inner Galaxy sources which leads to the near-far distance flips. These effects are confined to a minority of complex cases, demonstrating that the CO-based association method provides a robust, consistent, and homogeneous framework for determining kinematic distances in the entire sample of Galactic \ion{H}{ii} regions.

\begin{table*}
\centering
\caption{Comparing the ambiguous or multiple RRL velocity measurements from the WISE \ion{H}{ii} regions with the assigned 
CO-derived SMGPS velocities for the sources. From left to right are columns SMGPS extended \ion{H}{ii} region  name in IAU format, SMGPS source object name, associated SEDIGISM molecular cloud name, centroid velocity for the associated SEDIGISM molecular cloud, the cloud’s velocity range for the best-matching velocity window where all the associated clouds intersect, cloud's mean radial velocity, WISE \ion{H}{ii} region name, RRL velocity for the WISE \ion{H}{ii} region, the reference for the WISE RRL velocity measurement, and assigned CO-based velocity to the \ion{H}{ii} regions. The reference for all the velocities for the associated SEDIGISM molecular clouds (SDG $\mathrm{v_{lsr}}$ reference) is \protect\cite{2021MNRAS.Duarte}.}
\resizebox{\textwidth}{!}{
\begin{tabular}{ccccccccccc}
\hline
\hline
SMGPS & SMGPS & SEDIGISM & cloud v$_\text{lsr}$ & v$_\text{lsr}$ range & Mean SDG v$_\text{lsr}$ & WISE & WISE$_\text{VLSR}$ & WISE$_\text{VLSR}$ reference & Adopted SMGPS$_\text{SDG}$ vel\\
(\ion{H}{ii} region Name) & (objname) & (Cloud Name) & (\kms) & (\kms) & (\kms) & Name & (\kms) & - & (\kms) \\
\hline
G012.203-00.110 & G012.202-00.110 & SDG012.195-0.1109 & 25.45 & 19.00, 34.25 & 25.45  & G012.202-00.110 & 27.0; 53.6 & {\citealt{1989ApJS...Lockman}} & 25.45 \\
G012.209-00.104 & G012.208-00.106 & SDG012.195-0.1109 & 25.45 & 17.50, 32.25 & 25.45  & G012.208-00.106 & 27.0; 53.6 & {\citealt{1989ApJS...Lockman}} & 25.45 \\
G012.302-00.064* & G012.311-00.059 & SDG012.293-0.0819 & 65.40 & 61.75, 69.00 & 65.65  & G012.311-00.059 & 67.3; 30.5 & {\citealt{2011ApJS...Anderson}} & 65.65 \\
G012.302-00.064* & G012.311-00.059 & SDG012.285-0.0264 & 65.90 & 61.75, 69.00 & 65.65  & G012.311-00.059 & 67.3; 30.5 & {\citealt{2011ApJS...Anderson}} & 65.65 \\
G012.727-00.218 & G012.726-00.219 & SDG012.840-0.2041 & 35.48 & 31.75, 38.25 & 35.48  & G012.726-00.219 & 57.7; 33.8 & {\citealt{2015ApJ...Anderson_b}} & 35.48 \\
G013.323+00.093 & G013.322+00.096 & SDG013.322+0.0989 & 18.64 & 17.25, 20.25 & 18.64  & G013.322+00.096 & 12.3; 42.9 & {\citealt{2011ApJS...Anderson}} & 18.64 \\
G013.709-00.240 & G013.709-00.243 & SDG013.568-0.2492 & 18.24 & 15.75, 21.00 & 18.24  & G013.709-00.243 & 97.8; 36.3; 8.6 & {\citealt{2011ApJS...Anderson}} & 18.24 \\
G016.380-00.308 & G016.380-00.308 & SDG016.381-0.3126 & 42.23 & 42.00, 44.75 & 42.23  & G016.380-00.308 & 44.8; 14.5 & {\citealt{2011ApJS...Anderson}} & 42.23 \\
G016.416-00.371 & G016.419-00.371 & SDG016.430-0.3720 & 11.71 & 10.25, 13.25 & 11.71  & G016.419-00.371 & 10.1; 45.9; 95.8 & {\citealt{2011ApJS...Anderson}} & 11.71 \\
G312.594+00.224 & G312.591+00.210 & SDG312.605+0.2592 & -53.06 & -56.00, -50.00 & -53.06  & G312.591+00.210 & -63.2000; -33.0000 & {\citealt{2021ApJS..Wenger}} & -53.06 \\
G320.667+00.221* & G320.590+00.190 & SDG320.419+0.2039 & -69.30 & -77.50, -62.50 & -70.58  & G320.590+00.190 & -8.50000; -7.70000 & {\citealt{2021ApJS..Wenger}} & -70.58 \\
G320.667+00.221* & G320.590+00.190 & SDG320.604-0.0081 & -71.18 & -77.50, -62.50 & -70.58  & G320.590+00.190 & -8.50000; -7.70000 & {\citealt{2021ApJS..Wenger}} & -70.58 \\
G320.667+00.221* & G320.590+00.190 & SDG320.733-0.0362 & -67.20 & -77.50, -62.50 & -70.58  & G320.590+00.190 & -8.50000; -7.70000 & {\citealt{2021ApJS..Wenger}} & -70.58 \\
G320.667+00.221* & G320.590+00.190 & SDG320.824+0.2885 & -72.01 & -77.50, -62.50 & -70.58  & G320.590+00.190 & -8.50000; -7.70000 & {\citealt{2021ApJS..Wenger}} & -70.58 \\
G320.667+00.221* & G320.590+00.190 & SDG320.728+0.0020 & -69.93 & -77.50, -62.50 & -70.58  & G320.590+00.190 & -8.50000; -7.70000 & {\citealt{2021ApJS..Wenger}} & -70.58 \\
G320.667+00.221* & G320.590+00.190 & SDG320.828-0.0173 & -69.43 & -77.50, -62.50 & -70.58  & G320.590+00.190 & -8.50000; -7.70000 & {\citealt{2021ApJS..Wenger}} & -70.58 \\
G320.667+00.221* & G320.590+00.190 & SDG320.538+0.0606 & -74.97 & -77.50, -62.50 & -70.58  & G320.590+00.190 & -8.50000; -7.70000 & {\citealt{2021ApJS..Wenger}} & -70.58 \\
G321.030-00.504 & G321.015-00.527 & SDG321.016-0.4590 & -59.13 & -63.50, -55.00 & -59.13  & G321.015-00.527 & -51.7000; -71.7000 & {\citealt{2021ApJS..Wenger}} & -59.13 \\
G330.954-00.182 & G330.954-00.181 & SDG331.047-0.1960 & -90.05 & -100.00, -81.25 & -90.05  & G330.954-00.181 & -88.5000; -88.7000 & {\citealt{2021ApJS..Wenger}} & -90.05 \\
G332.139-00.439* & G332.145-00.452 & SDG332.290-0.3807 & -51.65 & -62.50, -50.00 & -54.39  & G332.145-00.452 & -55.6000; -80.0000 & {\citealt{2021ApJS..Wenger}} & -54.39 \\
G332.139-00.439* & G332.145-00.452 & SDG332.098-0.4255 & -56.97 & -62.50, -50.00 & -54.39  & G332.145-00.452 & -55.6000; -80.0000 & {\citealt{2021ApJS..Wenger}} & -54.39 \\
G332.139-00.439* & G332.145-00.452 & SDG332.080-0.4737 & -54.55 & -62.50, -50.00 & -54.39  & G332.145-00.452 & -55.6000; -80.0000 & {\citealt{2021ApJS..Wenger}} & -54.39 \\
G336.990-00.022 & G336.990-00.021 & SDG337.000-0.0138 & -118.47 & -125.75, -114.00 & -118.47  & G336.990-00.021 & -118.900; -96.0000 & {\citealt{2021ApJS..Wenger}} & -118.47 \\
G337.139+00.010 & G337.138+00.008 & SDG337.000-0.0138 & -118.47 & -117.00, -109.75 & -118.47  & G337.138+00.008 & -116.700; -68.6000 & {\citealt{2021ApJS..Wenger}} & -118.47 \\
G337.172-00.058 & G337.170-00.059 & SDG337.191-0.0461 & -70.06 & -70.00, -59.50 & -70.06  & G337.170-00.059 & -69.7000; -113.500 & {\citealt{2021ApJS..Wenger}} & -70.06 \\
G337.254-00.165 & G337.253-00.165 & SDG337.191-0.0461 & -70.06 & -65.25, -62.00 & -70.06  & G337.253-00.165 & -39.9000; -64.4000 & {\citealt{2021ApJS..Wenger}} & -70.06 \\
G337.286-00.113 & G337.285-00.113 & SDG337.191-0.0461 & -70.06 & -69.50, -61.75 & -70.06  & G337.285-00.113 & -42.2000; -41.8000 & {\citealt{2021ApJS..Wenger}} & -70.06 \\
G337.307-00.139 & G337.304-00.142 & SDG337.191-0.0461 & -70.06 & -73.25, -61.00 & -70.06  & G337.304-00.142 & -70.5000; -40.5000 & {\citealt{2021ApJS..Wenger}} & -70.06 \\
G337.710+00.091 & G337.709+00.091 & SDG337.684+0.1085 & -74.19 & -80.50, -70.75 & -74.19  & G337.709+00.091 & -78.5000; -44.8000 & {\citealt{2021ApJS..Wenger}} & -74.19 \\
G337.717+00.058* & G337.754+00.057 & SDG337.676+0.0823 & -67.28 & -68.75, -41.00 & -59.61  & G337.754+00.057 & -47.4000; -124.800 & {\citealt{2021ApJS..Wenger}} & -59.61 \\
G337.717+00.058* & G337.754+00.057 & SDG337.683-0.0424 & -51.95 & -68.75, -41.00 & -59.61  & G337.754+00.057 & -47.4000; -124.800 & {\citealt{2021ApJS..Wenger}} & -59.61 \\
G337.880+00.093 & G337.877+00.091 & SDG337.885+0.1121 & -31.15 & -31.50, -27.75 & -31.15  & G337.877+00.091 & -44.8000; -85.0000 & {\citealt{2021ApJS..Wenger}} & -31.15 \\
G338.002+00.080 & G337.996+00.081 & SDG337.998+0.0782 & -122.17 & -126.25, -118.25 & -122.17  & G337.996+00.081 & -129.900; -52.5000 & {\citealt{2021ApJS..Wenger}} & -122.17 \\
G339.233+00.226* & G339.233+00.243 & SDG339.297+0.3318 & -31.73 & -34.75, -23.50 & -30.02  & G339.233+00.243 & -70.6000; -30.0000 & {\citealt{2021ApJS..Wenger}} & -30.02 \\
G339.233+00.226* & G339.233+00.243 & SDG339.293+0.1184 & -28.32 & -34.75, -23.50 & -30.02  & G339.233+00.243 & -70.6000; -30.0000 & {\citealt{2021ApJS..Wenger}} & -30.02 \\
G342.115+00.028* & G342.120+00.001 & SDG342.151+0.0370 & -128.94 & -135.75, -123.00 & -129.46  & G342.120+00.001 & -131.0; -38.2 & {\citealt{2018ApJS..Anderson}} & -129.46 \\
G342.115+00.028* & G342.120+00.001 & SDG342.052+0.0505 & -129.98 & -135.75, -123.00 & -129.46  & G342.120+00.001 & -131.0; -38.2 & {\citealt{2018ApJS..Anderson}} & -129.46 \\
G343.350-00.330 & G343.344-00.330 & SDG343.312-0.3185 & -82.26 & -87.00, -76.50 & -82.26  & G343.344-00.330 & -82.3; -22.4; 1.0 & {\citealt{2018ApJS..Anderson}} & -82.26 \\
G343.352-00.067 & G343.353-00.068 & SDG343.340-0.0733 & -121.00 & -127.75, -115.50 & -121.00  & G343.353-00.068 & -130.500; -23.8000 & {\citealt{2021ApJS..Wenger}} & -121.00 \\
G343.641-00.168 & G343.636-00.150 & SDG343.627-0.1776 & -32.71 & -35.75, -29.75 & -32.71  & G343.636-00.150 & -26.5; -85.8 & {\citealt{2018ApJS..Anderson}} & -32.71 \\
G343.704-00.128 & G343.708-00.124 & SDG343.756-0.1620 & -26.11 & -28.75, -26.00 & -26.11  & G343.708-00.124 & -23.8; -84.9 & {\citealt{2015ApJ...Anderson_b}} & -26.11 \\
G343.717-00.215 & G343.712-00.210 & SDG343.756-0.1620 & -26.11 & -31.00, -23.25 & -26.11  & G343.712-00.210 & -26.8; -83.2 & {\citealt{2011ApJS...Anderson}} & -26.11 \\
G343.860-00.217 & G343.855-00.226 & SDG343.756-0.1620 & -26.11 & -28.25, -23.25 & -26.11  & G343.855-00.226 & -26.5; -86.2 & {\citealt{2011ApJS...Anderson}} & -26.11 \\
G346.054-00.020 & G346.056-00.020 & SDG346.063-0.0519 & -83.04 & -86.00, -75.00 & -83.04  & G346.056-00.020 & -76.8; -3.4 & {\citealt{2011ApJS...Anderson}} & -83.04 \\
G346.078-00.057 & G346.078-00.057 & SDG346.063-0.0519 & -83.04 & -87.50, -79.00 & -83.04  & G346.078-00.057 & -84.7; -7.9 & {\citealt{2011ApJS...Anderson}} & -83.04 \\
G348.148+00.255 & G348.148+00.256 & SDG348.137+0.2661 & -69.45 & -73.75, -65.25 & -69.45  & G348.148+00.256 & -66.3; -1.4 & {\citealt{2011ApJS...Anderson}} & -69.45 \\
G352.526-00.143 & G352.527-00.147 & SDG352.486-0.1722 & -50.28 & -55.75, -46.00 & -50.28  & G352.527-00.147 & -57.3; -38.0 & {\citealt{2011ApJS...Anderson}} & -50.28 \\
$\vdots$ & $\vdots$ & $\vdots$ & $\vdots$ & $\vdots$ & $\vdots$ & $\vdots$ & $\vdots$ & $\vdots$ & $\vdots$\\
\hline\hline
\end{tabular}
}
\begin{minipage}{18cm}
\vspace{0.2cm}
\footnotesize \textbf{Notes:}  Only SMGPS sources that associate with multiple SEDIGISM molecular clouds (at similar velocities) bear an asterisk (*) infront of them under column SMGPS (\ion{H}{ii} region Name). 
\end{minipage}                       
\label{tab:table2}
\end{table*}

\subsection{Physical properties of the molecular clouds}
\label{subsec: cloud's physical property and moment maps}

\ion{H}{ii} regions are significant in understanding massive star formation as they are formed due to the feedback of young massive protostars, and can potentially trigger new star formation. They associate with
massive stars in regions of ongoing stellar activity \citep[][]{2007ARA&A..Zinnecker}, creating shockwaves that compress the nearby gas and cause gravitational instability that eventually leads to the birth of new stars. In this effect, we further show an illustration of the physical relevance as a result of the associations between the SMGPS \ion{H}{ii} regions and the SEDIGISM molecular clouds and how the physical properties of these clouds further influence the effect of large scale star formation. 

\begin{figure*}
    \centering
    \includegraphics[width=\textwidth]{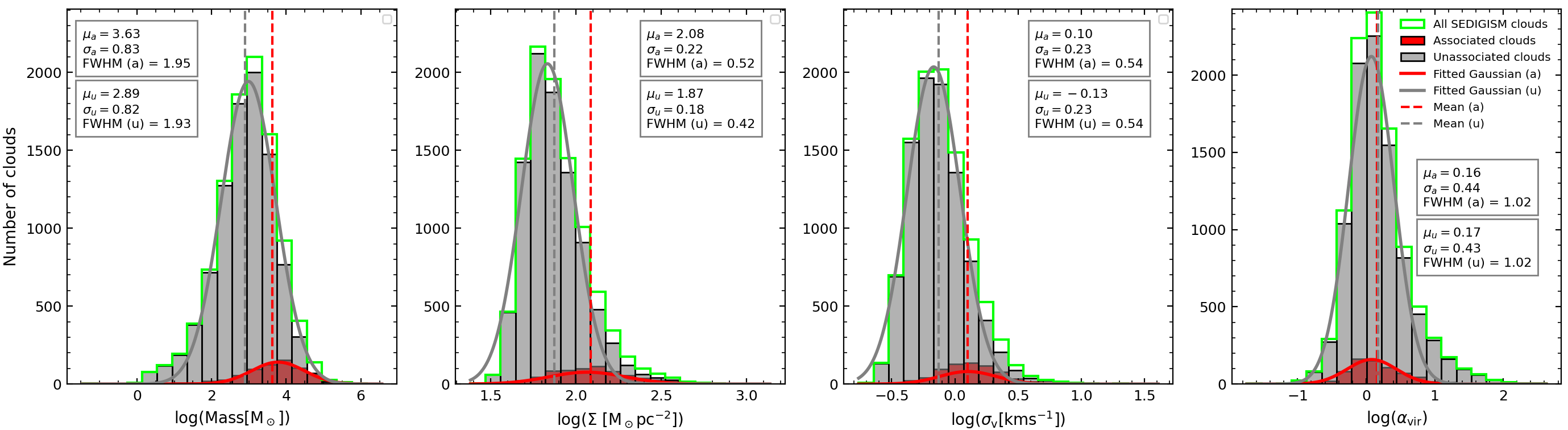}
    \caption{Histograms of key physical properties of both the associated clouds with the \ion{H}{ii} regions and the unassociated population. From left to right: Distribution of log of mass for the molecular clouds, distribution of log of deconvolved average gas surface density ($\Sigma$) of the clouds, distribution of log of clouds' velocity dispersion (linewidth) ($\sigma_\text{v}$), and distribution of log of deconvolved virial parameter($\alpha_\mathrm{vir}$) of the clouds, respectively. The Gaussian parameters are as shown and the bin size is 20 for each histogram. The values of mean mass, standard deviation, and Full Width at Half Maximum (FWHM) are as indicated with ${\textit{a}}$ and ${\textit{u}}$ representing associated and unassociated clouds, respectively. The third histogram in the back as a lime outline, shows the whole molecular cloud sample of the SEDIGISM survey.} 
    \label{fig:cloud_statistics}
\end{figure*}

Regarding the physical relevance and properties of associated and unassociated molecular clouds, in \cite{2026MNRAS...Langa},  we performed statistics of physical properties for the two cloud populations with SMGPS sources for only a single tile from each SMGPS and SEDIGISM survey. We have done the same here for the whole SEDIGISM survey (10663 clouds) to see what impact this brings about, specifically for the 684 associated clouds with the 741 \ion{H}{ii} regions. There are 9221 unassociated clouds. The remaining 758 clouds are associated with SNRs, unclassified sources, or other extended sources mentioned in Section \ref{sec: Data}. We have excluded such clouds from this statistics, as our primary interest in this paper is purely cloud associations with \ion{H}{ii} regions. Figure \ref{fig:cloud_statistics} presents histograms that compare, for the two categories of clouds, four cloud's physical properties (which we obtain directly from the SEDIGISM catalogue \citep{2021MNRAS.Duarte}): cloud's mass, average gas surface density, linewidth, and virial parameter. We observe that the associated clouds are more massive (exhibiting a significantly larger mean mass of 3.63 dex versus 2.89 dex for the unassociated clouds). In addition to mean mass, the associated clouds also possess a slightly wider distribution of masses as reflected in the Full Width at Half Maximums (FWHMs). We also observe a higher mean gas surface density for the associated clouds (2.08 dex) compared to the unassociated group (1.87 dex)-- an implication which is consistent with the expectation that massive star formation is more likely to occur in dense environments even though some associations are also found at lower surface densities. FWHM of the average gas surface density also shows a slight spread in distribution of the physical property in favour of the associated cloud population. Similarly, the distribution of linewidths (velocity dispersions, $\sigma_v$) shows that the associated clouds have a higher mean linewidth (0.10 dex) than the unassociated clouds (-0.13 dex). This suggests that clouds associated with the SMGPS \ion{H}{ii} regions tend to have larger internal turbulent motions. However, the FWHM values for the two cloud populations show that both exhibit equal spread of linewidth.  Although the mean virial parameter ($\alpha_\text{vir}$) for the unassociated clouds slightly edges the associated clouds', the distributions of virial parameters, however, show much less separation between the two population of clouds. We refer the reader to \cite{2026MNRAS...Langa} for a more in-depth discussion on the virial parameter.

\begin{figure}
    \centering
    \includegraphics[width=\columnwidth]{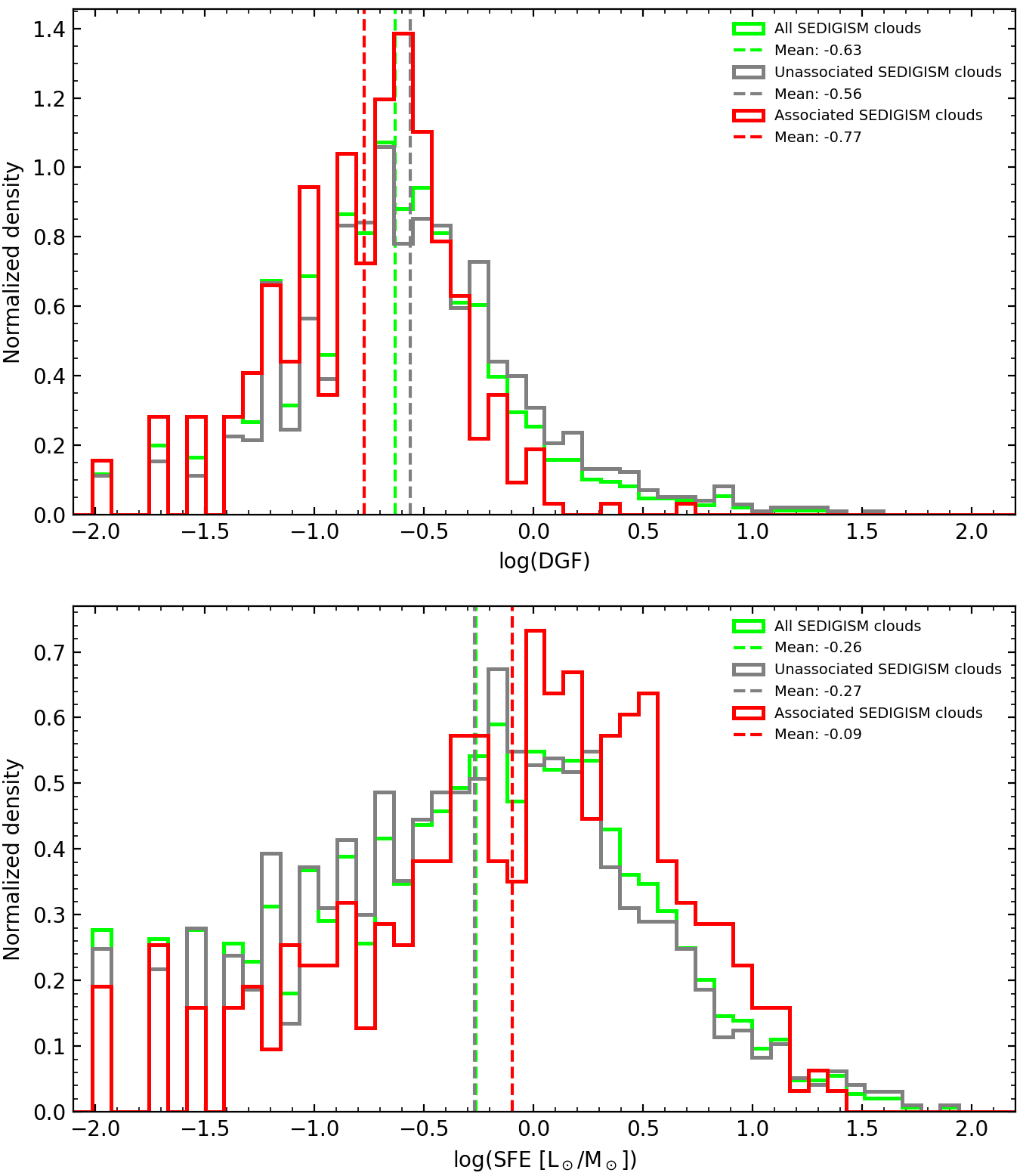}
    \caption{Top panel: Histogram of log of DGF of the whole SEDIGISM molecular clouds in the field in lime, unassociated SEDIGISM clouds in gray and associated SEDIGISM clouds with the \ion{H}{ii} regions in red. Bottom panel: Histogram of log of SFE (L$_\odot$/M$_\odot$) of the whole SEDIGISM clouds in the field in lime, unassociated clouds in gray and associated clouds in red. Also shown are the legends of the mean for the DGF and SFE (L$_\odot$/M$_\odot$) for the whole SEDIGISM survey, unassociated clouds and of the associated SEDIGISM molecular clouds. Note that in our previous study \citep{2026MNRAS...Langa} we explained that a small fraction (152/10663) of the SEDIGISM clouds exhibit DGF > 1 (log DGF > 0) as this arises from the large intrinsic uncertainties in the cloud and clump mass estimates, both of which were normalised to a common distance \citep{2018MNRAS...Urquhart, 2021MNRAS...Urquhart}.}
    \label{fig:SFE_DGF_mean}
\end{figure}

In addition, we further present the molecular clouds' statistics on star formation efficiency (SFE), dense gas fraction (DGF), and the presence or absence of a generic high-mass star formation (HMSF) indicator. Here, we maintain the definitions of these properties as in \citet{2013MNRAS...Urquhart_a,2014MNRAS...Urquhart, 2021MNRAS...Urquhart, 2021MNRAS.Duarte} and as further clarified in \cite{2026MNRAS...Langa}.
In summary and on the basis of this particular analysis, SFE$_\mathrm{GMC}$ is defined as the ratio of the total bolometric luminosity of ATLASGAL clumps within a giant molecular cloud (GMC) to the total mass of that GMC (\lclump/\mgmc)-- a measure of the instantaneous star formation efficiency. We also use DGF$_\mathrm{GMC}$ as the fraction of the GMC mass traced by compact dust emission, i.e. \mclump/\mgmc. \cite{2021MNRAS...Urquhart} provide SFE$_\mathrm{GMC}$ and DGF$_\mathrm{GMC}$ consistently for the SEDIGISM GMC sample, and we therefore apply the measurements directly in the statistics. Figure \ref{fig:SFE_DGF_mean} illustrates the distributions of SFE and DGF of the entire SEDIGISM clouds, associated clouds and unassociated clouds. \cite{2021MNRAS...Urquhart} reveal that there is only 22 per cent of all the SEDIGISM clouds with valid SFE and DGF measurements, hence referred to as having an ATLASGAL counterpart. Therefore, from the distributions, 365 clouds of the 684 associated clouds ($\sim$53 per cent) have both SFE and DGF measurements. The values clearly suggest higher fraction as compared to the entire SEDIGISM clouds with the same properties in the field. In contrast, only 1122 of the 9221 ($\sim$12 per cent) from the unassociated clouds have SFE and DGF measurements--fractions way lower than the ones from the associated clouds and even all the clouds. Generally, this is a clear indication that the clouds with associated SMGPS \ion{H}{ii} regions are more likely to be dense, actively star-forming molecular environments and therefore associated with ATLASGAL clumps. 

On the other hand, HMSF is classified by \cite{2021MNRAS.Duarte} who flagged SEDIGISM molecular clouds as having a HMSF tracer (=1) or not (=0), as per \cite{2014MNRAS...Urquhart}. We note that the SEDIGISM catalogue identified high-mass star formation (HMSF) tracers based purely on cross-matches with ATLASGAL sources, but not direct from the HMSF tracers \citep{2021MNRAS.Duarte}. The catalogue registers only 435 SEDIGISM clouds ($\sim$4 per cent of the full cloud sample) as having signposts of active HMSF. However, this work of identifying more SEDIGISM clouds with active HMSF is still being pursued as it depends on the sensitivity and coverage of the dust continuum data and is explicitly limited to ultracompact \ion{H}{ii} regions due to the use of the \citet{2014MNRAS...Urquhart} sample. Therefore our analysis from the SEDIGISM cloud-SMGPS \ion{H}{ii} region association provides a more complete tracer of ongoing high-mass star formation by extending the cross-matching to larger classical \ion{H}{ii} regions. 169 out of 684 associated SEDIGISM molecular clouds are already identified as HMSF tracers in the SEDIGISM catalogue. Therefore our cross-match offers a new sample of 515 molecular clouds with active HMSF that were not previously flagged as HMSF in the SEDIGISM catalogue. The inclusion of additional HMSF sample will significantly expand the known population of molecular clouds directly linked with active high-mass star formation, highlighting the importance of combining molecular line and radio continuum surveys.

\subsection{Ionising photon flux and Galactic environment}
\label{subsec: photon flux and distance}

\begin{figure}
    \centering
    \includegraphics[width=\columnwidth]{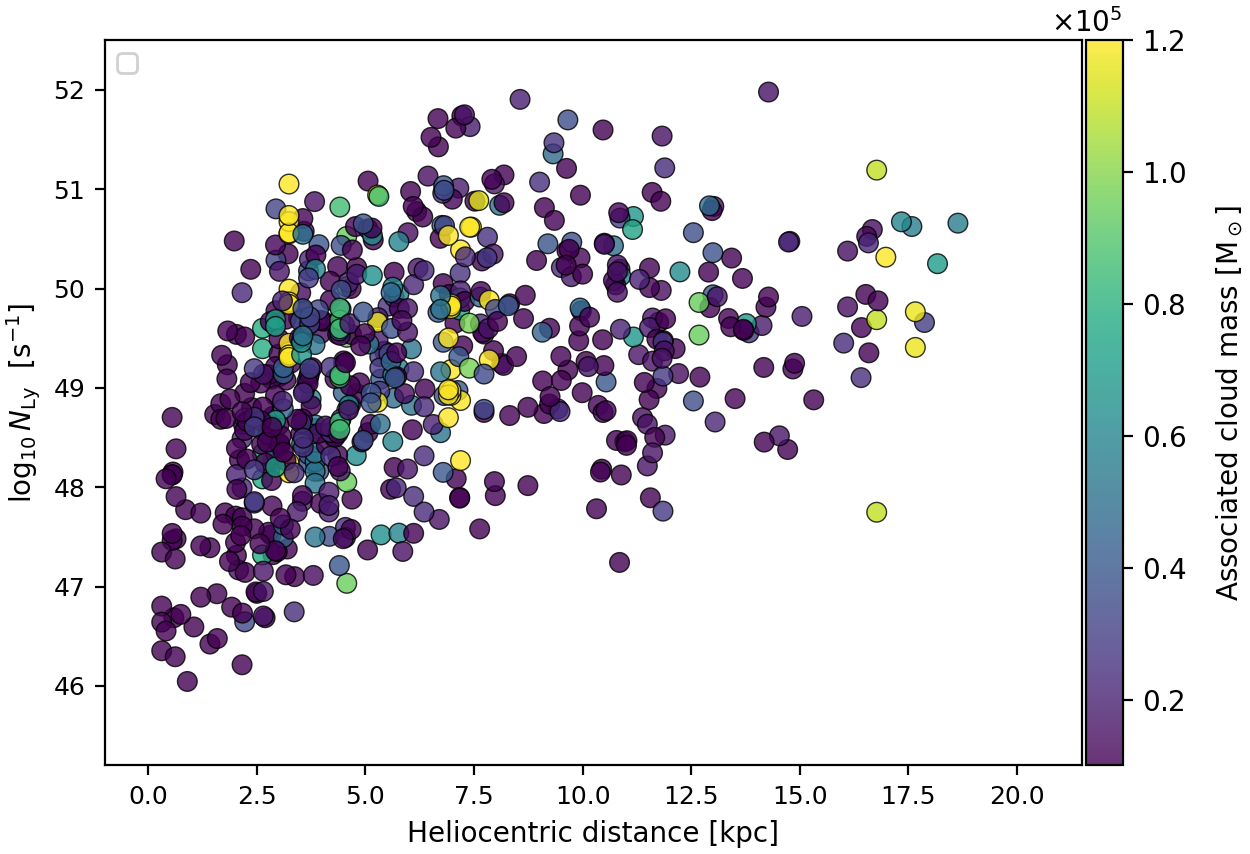}
    \caption{Ionising photon flux of the SMGPS \ion{H}{ii} regions as a function of adopted CO-based distance. The photon rates are derived from the 1.3 GHz radio continuum flux densities \citep[SMGPS catalogue;][]{2025A&A...Bordiu} assuming optically thin free–free emission and an electron temperature of $10^4$K. Points are colour-coded by the mass of the associated SEDIGISM molecular cloud.}
    \label{fig: Photon_flux_vs_distance}
\end{figure}

In Figure \ref{fig: Photon_flux_vs_distance} we present the ionizing photon flux of the 640 \ion{H}{ii} regions, as a function of the adopted CO-based distance for our sample, with points color-coded by the mass of the associated SEDIGISM molecular cloud. As a requirement to maintain an optically thin and ionization-bounded \ion{H}{ii } region, we have calculated the radio continuum ionizing photon flux following \citet{1967ApJ...Mezger, 1968ApJ...Rubin, 1994ApJS...Kurtz, 2018A&A...Kalcheva}:

\begin{equation}
N_{\mathrm{Ly}} \approx 4.761 \times 10^{48} \  
\left[ \frac{S_{\nu}}{\mathrm{Jy}} \right]
\left[ \frac{D}{\mathrm{kpc}} \right]^2
\left[ \frac{\nu}{\mathrm{GHz}} \right]^{0.1}
\left[ \frac{T_{\mathrm{e}}}{10^4 \ \mathrm{K}} \right]^{-0.45} \ \mathrm{s}^{-1}, 
\label{eq:photon flux}
\end{equation}
where $S_{\nu}$ is the integrated flux density of the radio source \citep[SMGPS catalogue;][]{2025A&A...Bordiu} at frequency $\nu$, $D$ is the adopted CO-based distance, and $T_{\mathrm{e}}$ is the electron temperature. For our analysis, we adopted $\nu = 1.3$ GHz (the SMGPS observing frequency) and $T_{\mathrm{e}} \approx 10^4$ K--a typical value for Galactic \ion{H}{ii} regions \citep{2014ApJS..Anderson}.

The distribution reveals that the most luminous \ion{H}{ii} regions ($N_{\mathrm{Ly}}>10^{49}\ \mathrm{s}^{-1}$) are predominantly detected at distances beyond $\sim$3 kpc, with their numbers peaking between $\sim$4 and 9 kpc. As expected, it also appears that there is a positive correlation between cloud mass and ionizing photon flux, since the most massive molecular clouds ($>10^5\ \mathrm{M_\odot}$, yellow points) are preferentially associated with the most luminous \ion{H}{ii} regions, consistent with expectations that large scale star formation is supported in the most massive molecular complexes \citep{2014prpl.Dobbs, 2018MNRAS...Urquhart}. We also observe that lower-mass clouds (purple points) host \ion{H}{ii} regions across the full range of luminosities but are predominantly found at closer distances where our sensitivity to faint emission is greater. There is a lack of correlation with distance, indicating that the adopted CO-derived distances yield physically consistent ionising photon estimates.

\begin{figure}
    \centering
    \includegraphics[width=\columnwidth]{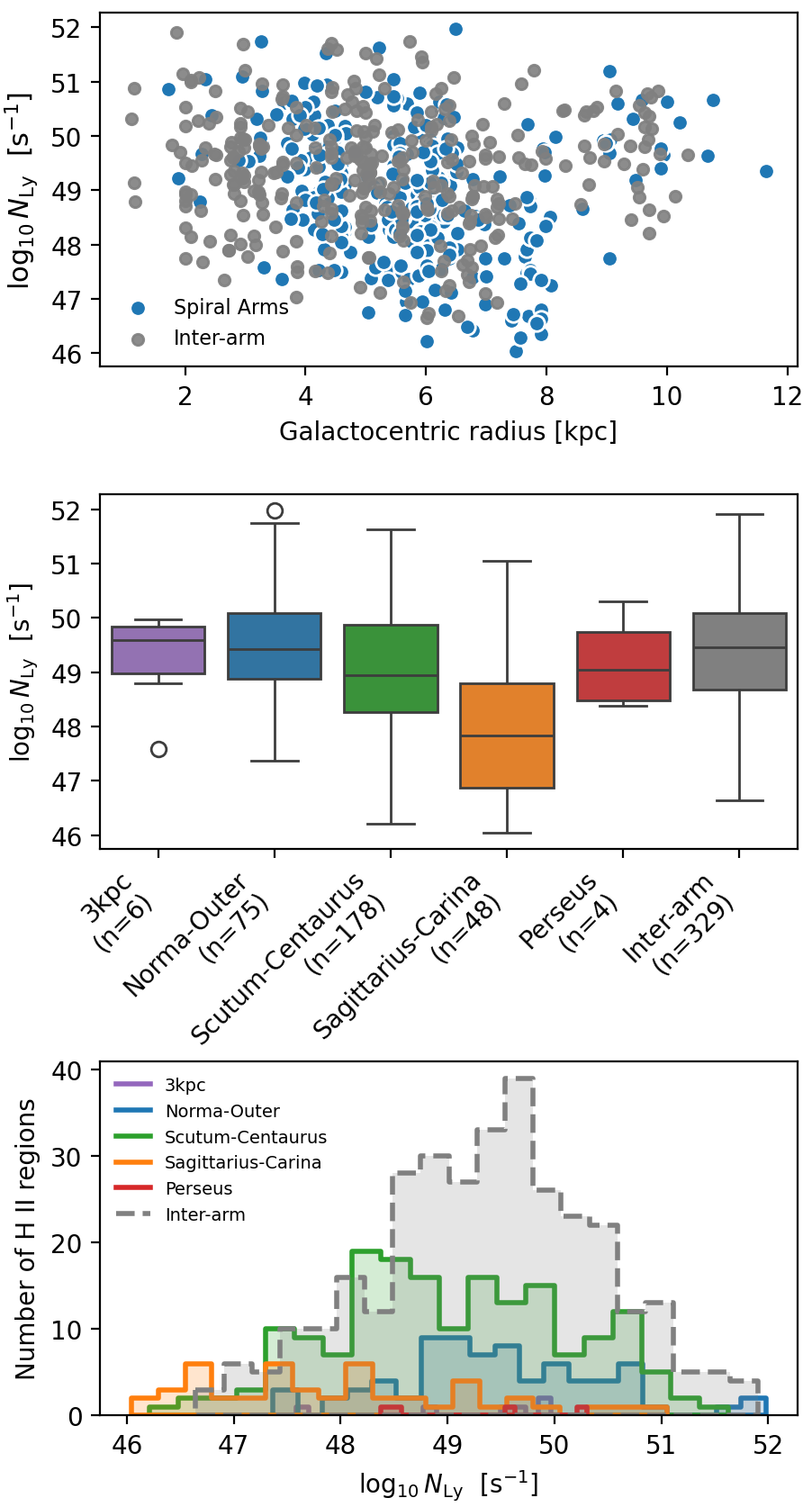}
    \caption{Top panel: Distribution of ionising photon flux of 640 SMGPS sources (blue--311 in spiral arms and gray--329 in inter-arm regions) as a function of Galactocentric radius. Middle panel: Boxplot distribution of ionising photon flux versus sources in each spiral arm and sources in inter-arm regions. Bottom panel: Histogram of ionising photon flux as a function of density of sources in both spiral arms and inter-arms.}
    \label{fig: ionising photos_n_multiple panels}
\end{figure}

We also investigate how the ionising photon flux ($N_\mathrm{Ly}$) varies with Galactic environment, including Galactocentric radius, spiral arm association, and cloud properties. In Figure \ref{fig: ionising photos_n_multiple panels} (top panel), we present the distribution of ionising photon flux of the \ion{H}{ii} regions as a function of Galactocentric radius, with the \ion{H}{ii} regions colour-coded according to whether they are associated with spiral arm or inter-arm molecular clouds. We observe a broad scatter in the alignment of flux sample, with both the spiral arm and inter-arm environments hosting \ion{H}{ii} regions spanning $\log_{10}N_\mathrm{Ly}\approx 46-52\ \ \mathrm{s}^{-1}$ across the entire radial range ($\sim2-12$\,kpc), but without any clear systematic Galactic trends. Likewise, the most luminous regions are found in both spiral arm and inter-arm populations, an indication that the luminosity of \ion{H}{ii} regions perhaps is a reflection of local conditions within individual molecular clouds but does not strongly depend on Galactocentric position or large scale Galactic environment. This trend agrees with some of the previous studies of compact and ultra-compact \ion{H}{ii} regions, which also find large intrinsic scatter and no strict correlation \citep[e.g.][]{2003A&A...Martin, 2024A&A...Dey}. 

\begin{figure}
    \centering
    \includegraphics[width=\columnwidth]{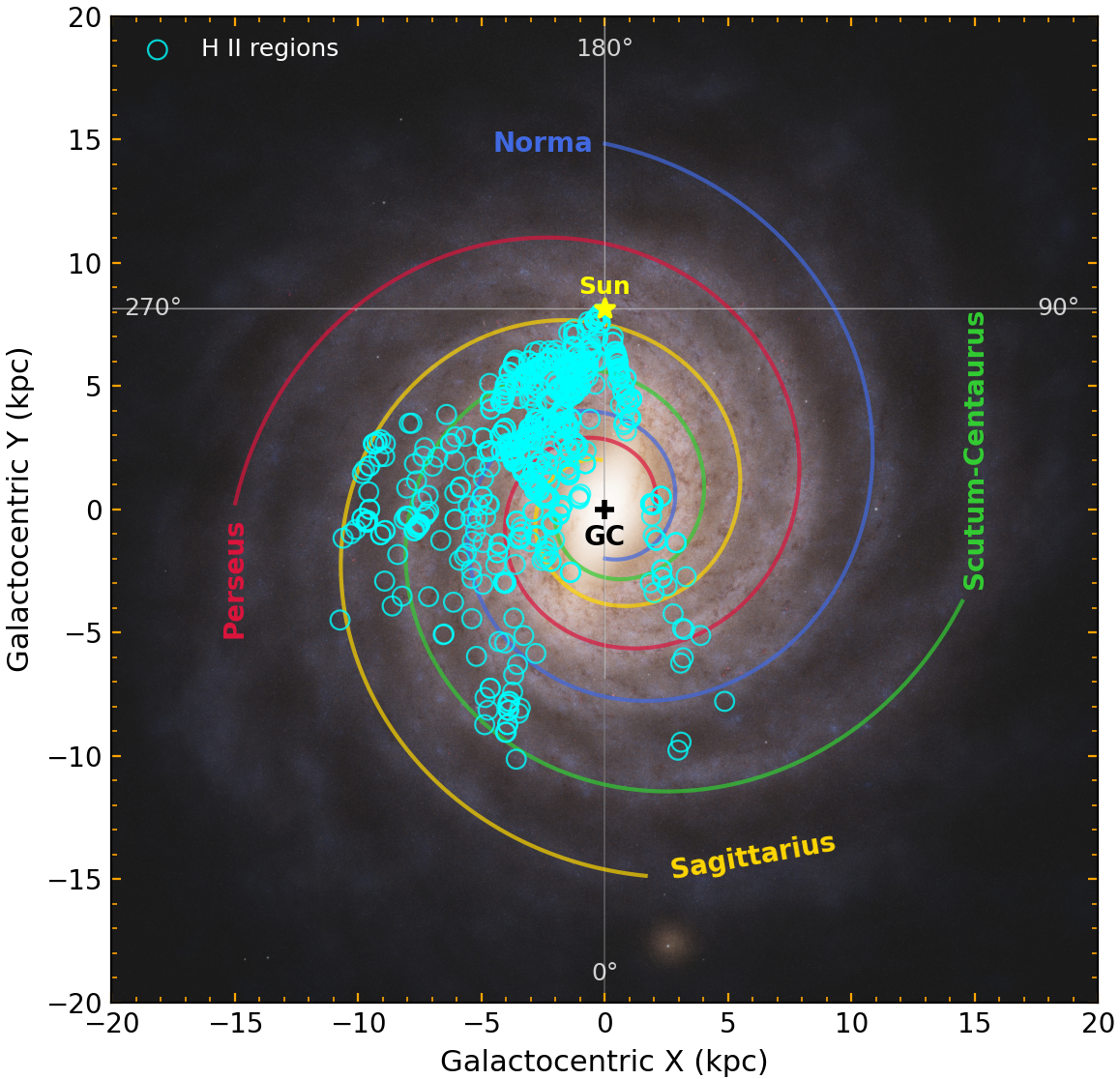}
    \caption{Galactocentric distribution of the 640 SMGPS \ion{H}{ii} regions (cyan circles) overlaid on the Milky Way background in a top-down view of the image of the Galaxy by The European Space Agency’s Milky Way-mapper Gaia (also see The European Space Agency (ESA) website\protect\footnote{\url{https://www.esa.int/Science_Exploration/Space_Science/Gaia/Last_starlight_for_ground-breaking_Gaia}} for more details). The Galactic Centre (GC) and the position of the Sun are marked in black cross and yellow star, respectively, with the main spiral arms are traced for reference.}
    \label{fig:Galactocentric_HII_distribution}
\end{figure}

To explicitly quantify the environmental effects on massive star formation, we further investigate the distribution of the SMGPS \ion{H}{ii} regions within and outside spiral arms, and in different spiral arms as well, using the spiral arm classifications provided in the SEDIGISM catalogue. The distribution across the arms is as in the middle and lower panels of Figure \ref{fig: ionising photos_n_multiple panels}. Among the sources associated with spiral arms, Scutum-Centaurus hosts the majority of the \ion{H}{ii} regions (178), followed by Norma (75), Sagittarius-Carina (48), the 3kpc arm (6), and Perseus (4), giving a total of 311/640 ($\sim$49 per cent) \ion{H}{ii} regions that are associated with 274/572 ($\sim$48 per cent) molecular clouds that lie within the spiral arms. In addition, the catalogue identifies a substantial inter-arm population consisting of 329/640 ($\sim$51 per cent) \ion{H}{ii} regions associated with 298/572 ($\sim$52 per cent) molecular clouds occupying the inter-arm regions. This demonstrates that while spiral arms are believed to be the most prolific stellar nurseries in the Galaxy, containing the highest concentration of dense molecular gas, they are not the only place stars are born--perhaps a good fraction of massive star formation also occurs outside the main spiral arm loci. We also examine $N_\mathrm{Ly}$ by spiral-arm environment and by individual spiral arms (see Figure \ref{fig: ionising photos_n_multiple panels}-middle panel), where we observe slight variations in the median luminosities across the different populations, showing slight enhancement in some particular structures--the ionising luminosity distribution tends to slightly vary with Galactic environment. The inter-arm population exhibits a median ionising luminosity comparable to those of the 3kpc, Norma, and Perseus arms, indicating that \ion{H}{ii} regions outside spiral arms are not systematically less luminous than those within them. Among the spiral arms, the 3kpc (only 6 sources), Norma, Scutum-Centaurus, and Perseus (although it has only 4 sources) arms exhibit a tendency towards higher luminosities, whereas the Sagittarius-Carina arm is characterised by systematically lower ionising luminosities. In particular, a Kolmogorov-Smirnov (K-S) test comparing Sagittarius-Carina against all remaining spiral arms reveals that $N_\mathrm{Ly}$ distribution yields a $p$-value of $6.31\times 10^{-9}$, allowing us to reject the null hypothesis and conclude that the Sagittarius-Carina population is statistically significant and characterised by systematically lower ionising luminosities. Perhaps this could be due to the viewing angle of SEDIGISM, for instance, in Figure \ref{fig:Galactocentric_HII_distribution} we can observe that the part of Sagittarius-Carina that is nearest to the Sun is at a Galactic longitude that is outside the SEDIGISM limit of $\ell=300^{\circ}$. Furthermore, comparison with the inter-arm population also reveals a significant difference ($p$-value of $3.00\times 10^{-11}$), indicating that the lower luminosities observed in Sagittarius-Carina are not representative of inter-arm star formation in general. However, the distributions generally indicate that spiral arms do not preferentially host significantly more luminous regions. The higher median luminosities in the mentioned spiral arms and inter-arm regions may suggest potentially more powerful star formation events in these Galactic regions. 

We further tested whether the Star Formation Efficiency (SFE) is enhanced in specific environments. Considering only molecular clouds within the spiral arms, the Sagittarius-Carina arm exhibits the highest mean SFE (3.47 $\mathrm{L_\odot/M_\odot}$), followed by Norma (2.57 $\mathrm{L_\odot/M_\odot}$), Scutum-Centaurus (1.62 $\mathrm{L_\odot/M_\odot}$), the 3kpc arm (1.53 $\mathrm{L_\odot/M_\odot}$), and Perseus (0.41 $\mathrm{L_\odot/M_\odot}$, although only a single cloud has an SFE measurement). Comparing the spiral-arm and inter-arm populations, the mean SFE is 2.14 and 2.41 $\mathrm{L_\odot/M_\odot}$, respectively. However, a K-S test yields K-S = 0.101 and $p = 0.374$, indicating no statistically significant difference between the two distributions. Thus, within our sample, we find no evidence that star formation efficiency is systematically enhanced in molecular clouds lying within the spiral arms relative to clouds in inter-arm regions.

Finally, we examine the cloud morphological types \citep[as per the][]{2022A&A...Neralwar} of the associated SEDIGISM molecular clouds to address whether specific cloud structures preferentially host \ion{H}{ii} regions. From the spiral arm environment, we find that elongated clouds are the most common host morphology, accounting for $\sim$42.7 per cent of the associations. Ring-like structures, which are often indicative of feedback-driven expansion, represent the second most common morphology at $\sim$26.6 per cent, followed by concentrated ($\sim$14.6 per cent) and clumped ($\sim$13.5 per cent) clouds. On the other hand, the inter-arm regions are also dominated by elongated clouds (38.6 per cent), followed by ring structures (23.1 per cent), clumped clouds (20.1 per cent), and concentrated structures (15.4 per cent). Comparing the catalogue-defined spiral-arm and inter-arm populations, the fraction of ring-like clouds is $\sim$26.6 per cent in spiral arms and $\sim$23.2 per cent in inter-arm environments; a fraction slightly higher in favour of spiral arm environment. However,  a chi-square test yields $p = 0.280$, indicating no statistically significant dependence of cloud morphology on spiral-arm environment. This diverse morphological breakdown indicates that while expanding, ring-like bubbles are significant, the majority of massive star formation in our sample is still heavily embedded in elongated filamentary molecular cloud networks irrespective of whether the clouds lie in spiral-arm or inter-arm environments.

\section{Discussion}
\label{sec: Discussion}

\subsection{The Galactic distribution of \ion{H}{ii} regions}
\label{subsec: HII_Galactic distribution}

We strongly believe that the reliable kinematic distances derived via our association method enable us to trace the large-scale distribution of massive star formation in the Galactic plane. To confirm our argument and analysis, we interpret Figure \ref{fig:Galactocentric_HII_distribution}, which presents the Galactocentric (X,Y) positions of the 640 SMGPS \ion{H}{ii} regions (cyan circles). The X and Y Galactic coordinates are directly adopted to our work as provided in the SEDIGISM catalogue \citep{2021MNRAS.Duarte}. As such, the kinematic distances for the associated SEDIGISM clouds were originally estimated using the Galactic rotation curve model \citep{2016ApJ...Reid} which assumed $R_0 = 8.34$ kpc, the distance to the Sun from the Galactic center, and $V_0 = 240$\,\kms for the Solar circular rotation speed. Although these are older values for the kinematic estimates as compared to the updated values \citep[e.g.,  $R_0 = 8.15$ kpc;][]{2019ApJ...Reid}, but still they give results that are within the Galactic typical margin of error for the kinematic distance method. This map offers a wide view of the spatial distribution of massive star-forming regions across the inner Milky Way.

We have observed that the resulting distribution is clearly not random. Therefore, to explicitly test the relationship between massive star formation and Galactic structure, we have done a statistical analysis on the radio continuum sources relative to the major spiral arms. Specifically, according to the spiral-arm classification provided in the SEDIGISM catalogue, 311 of the 640 associated SMGPS \ion{H}{ii} regions ($\sim$49 per cent) are associated with spiral-arm molecular clouds, while 329 ($\sim$51 per cent) are associated with inter-arm molecular clouds. In particular, the distribution in Figure \ref{fig: ionising photos_n_multiple panels} has revealed that, among the spiral arm distribution, the Scutum-Centaurus arm hosts the vast majority of \ion{H}{ii} regions (178), followed by Norma (75), Sagittarius-Carina (48), the 3kpc arm (6), and Perseus (4). Spiral arms are believed to be crucial sites of massive star formation \citep{2015ApJ...Nguyen}, where gas compression, cloud clustering, and star formation efficiencies are actually enhanced \citep{2014prpl.Dobbs}. However, the Galactic distribution of clouds with SMGPS \ion{H}{ii} regions (Figure \ref{fig:Galactocentric_HII_distribution}) suggests that these are not always aligned with segments of the Galaxy's major spiral arms in a top-down view of the Galaxy. Indeed, these results would instead suggest that although spiral arms host molecular gas concentration and massive star formation, they are not the exclusive locations where massive stars are formed. The presence of a substantial inter-arm population demonstrates that massive star formation is not confined exclusively to the main spiral-arm loci, but also occurs throughout the inter-arm molecular cloud population identified by the SEDIGISM catalogue.

We further examine the distribution of ionising photon flux across the Galaxy. In Figure \ref{fig: Photon_flux_vs_distance}, we have shown that there is lack of correlation between the ionizing photon flux and distance, which indicates that the adopted CO-based distances for the \ion{H}{ii} regions are robust, and further reinforces the physical association between the radio continuum sources and their parent molecular clouds. The ionizing photon rates indicate that the SMGPS extended \ion{H}{ii} regions spread across the Galactic arm environments for massive star formation with the most luminous \ion{H}{ii} regions are associated with the most massive molecular clouds. In addition, in Figure \ref{fig: ionising photos_n_multiple panels},
the results show no strong dependence of luminosity on Galactocentric radius or spiral arm location, with a large intrinsic scatter present in both spiral-arm and inter-arm environments. The substantial overlap between the two populations indicates that \ion{H}{ii} regions in inter-arm molecular clouds are not systematically less luminous than those associated with spiral arms, suggesting that the formation of massive stars is mainly determined by local cloud conditions rather than global Galactic structure. Similar conclusions on the scatter have been reported in studies of Galactic distribution of \ion{H}{ii} regions (especially compact and ultra-compact \ion{H}{ii} regions)\citep[e.g.][]{2003A&A...Martin, 2004MNRAS....Paladini, 2017ApJ....Makai, 2024A&A...Dey}, where luminosity distributions show weak or absent large-scale gradients. Although the Sagittarius-Carina arm exhibits systematically lower ionising luminosities than the remaining spiral arms, the overall comparison between spiral-arm and inter-arm populations reveals no evidence that \ion{H}{ii} regions within spiral arms are systematically more luminous. Furthermore, investigations into SFE across the Galactic disk give no statistically significant difference between the spiral-arm and inter-arm molecular clouds, indicating that the large number of \ion{H}{ii} regions associated with spiral arms is more likely a consequence of the concentration of molecular gas rather than a fundamental increase or intrinsic enhancement in the star-forming efficiency of individual clouds. Overall, our results therefore support the interpretation of \cite{2022A&A...Colombo}, in which spiral arms primarily act as sites of molecular gas concentration or accumulation rather than fundamentally enhancing star formation efficiency.

Beyond tracing the spiral structure, regions of high source density and potential line-of-sight crowding are also revealed in the map, specifically, toward the tangent points of spiral arms to the Sun's position \citep{2014A&A...Hou}. Therefore, it is believed that it is in these crowded environments that traditional kinematic distance methods are most prone to confusion due to velocity crowding and near–far ambiguities. In that connection, and with every effort put, even some of the kinematic distances for the SEDIGISM clouds still suffered from the need to solve the KDA (especially those flagged $d_{\mathrm{flag}}=12$), and some remain uncertain and unreliable ($d_\text{reliable}=0$) , and that includes the Galactic center clouds. Therefore, although our cloud-association technique provides a robust, physically motivated method for assigning systemic velocities to SMGPS \ion{H}{ii} regions, it fundamentally inherits the distance uncertainties and unresolved near-far ambiguities of the host molecular clouds. Consequently, the spatial pile-ups observed at the tangent points and near the Galactic Centre reflect the persistent challenges of the kinematic distance method in these complex regions.

\subsection{\ion{H}{ii} regions: Implications of massive star formation}

The SMGPS–WISE–SEDIGISM associations we have established in the study provide us with a physically grounded framework to enable us to examine large scale star formation and feedback in the inner Milky Way. By associating radio continuum \ion{H}{ii} regions to their parental molecular clouds \citep{2021MNRAS...Urquhart, 2021MNRAS.Duarte, 2022A&A...Neralwar, 2022A&A...Neralwar_b} through velocity coherence and spatial overlap, we directly link sites of massive stellar activity to the molecular environments from which they formed.

As illustrated in the previous sections, a major and central outcome of our study is the assignment of robust systemic velocities and kinematic distances to a large, homogeneous sample of WISE \ion{H}{ii} regions-including many radio continuum sources lacking RRL measurements or exhibiting ambiguous or multiple RRL velocities. Consequently, because the SEDIGISM molecular clouds trace the dense molecular gas associated with massive star formation \citep[e.g.,][]{2017A&A...Schuller, 2021MNRAS..Schuller, 2021MNRAS...Urquhart, 2021MNRAS.Duarte, 2022A&A...Neralwar, 2022A&A...Neralwar_b}, the resulting kinematic distances are physically motivated and less affected by line-of-sight confusion than we have noticed in the cases with only RRL assignments. Therefore, this substantially supports and improves the completeness and reliability of the Galactic massive star formation survey, particularly in the inner Galaxy where velocity crowding is seen to be severe. The strong correspondence between CO cloud velocities and WISE RRL (ionised gas) velocities in Figure \ref{fig:SMGPS_WISE}, together with the predominance of one-to-one cloud–\ion{H}{ii} region associations ($\sim$84 per cent; see Figure \ref{fig: histogram_clouds_HII}), further confirms that molecular gas provides a reliable tracer of the radial velocity of \ion{H}{ii} regions. This supports the expectation that, in the absence of extreme feedback-driven decoupling, ionised and molecular components remain dynamically linked. There are some exceptions, however, with large velocity discrepancies (>30–50 \kms) and we cautiously interpret these as arising from line-of-sight confusion, multiple embedded regions within a single beam, or evolved systems where feedback has significantly altered the local gas kinematics.

Massive star formation requires intensive physical involvement and therefore the emission fraction metric, \fw, \citep{2026MNRAS...Langa} introduced in this work also provides a quantitative measure of the physical interaction between the \ion{H}{ii} regions and the associated molecular clouds. Thus, high \fw\, values identify systems where the ionised emission is tightly embedded and has much brightness within a dominant molecular structure, while lower \fw\, values indicate weaker or ambiguous associations with lower emission intensity (see Section \ref{subsec: velocity_validation}). Additionally, the emission fraction allows us to resolve multiple velocity ambiguities in a systematic way, selecting only the cloud most likely to represent the natal environment of the \ion{H}{ii} region. These kinematic signatures are consistent with ongoing interactions between the ionised and molecular components within the host clouds, rendering them gravitationally unstable, and leading them to collapse to form a new generation of stars \citep{2009A&A...Pomares}, although the present analysis does not directly constrain whether feedback is triggering subsequent generations of star formation. The cause to massive star formation is further reinforced by the physical properties of the associated SEDIGISM molecular clouds. In Figures \ref{fig:cloud_statistics} and \ref{fig:SFE_DGF_mean}, we have observed that clouds linked to \ion{H}{ii} regions frequently exhibit high masses (up to $\sim10^5\,\text{M}_\odot$), elevated average gas surface densities, dense gas fraction (DGF) measurements, and low virial parameters. The presence of SMGPS \ion{H}{ii} regions further confirms that these molecular clouds are actively forming high-mass stars and the corresponding distributions of SFE and DGF indicate that the radio continuum sources are embedded within dense and dynamically evolved molecular environments, consistent with active and potentially ongoing star formation.

Perhaps the result with the most implications, beyond the kinematic associations, emerging from our analysis is the strong association between the most luminous \ion{H}{ii} regions and the most massive molecular clouds, as observed in Figure \ref{fig: Photon_flux_vs_distance}, where sources with the highest ionising photon fluxes ($N_\text{Ly}$) are preferentially linked to or predominantly embedded within SEDIGISM molecular clouds with the largest molecular gas reservoirs, often exceeding $\sim10^5\,\text{M}_\odot$. This is related to the mass reservoirs required for massive cluster formation. Usually, in the case where the \ion{H}{ii} regions cleared out their natal material as they evolved, it could be expected that the most luminous and powerful \ion{H}{ii} regions would be surrounded by less molecular gas, having already evacuated their environments. Instead, our findings suggest the opposite: these systems still retain substantial amounts of molecular material despite already hosting massive ionising stars. This suggests that there is a larger reservoir of mass around not just by chance and the formation of luminous \ion{H}{ii} regions does not rapidly disperse the parental cloud on GMC scales, but instead occurs within environments where the large reservoir of dense gas remains available. It implies that the surrounding cloud-scale environment plays a fundamental role in regulating the formation of massive stellar clusters, with the availability of large molecular gas reservoirs likely being a necessary condition for the formation of the most luminous and massive \ion{H}{ii} regions. Additionally, these highly luminous sources are confined within major spiral arms of which most associated clouds are elongated, ring-like, and clumpy in terms of morphological structures, evidence indicating higher massive star formation activity as compared to the other morphological types \citep{2022A&A...Neralwar, 2022A&A...Neralwar_b}. Overall, similar conclusions have been reported in Galactic studies showing that massive star formation preferentially occurs within the most massive and dense molecular cloud complexes \citep[e.g.][]{2014MNRAS...Urquhart, 2018MNRAS...Urquhart, 2021MNRAS...Urquhart, 2021MNRAS.Duarte}. This interpretation is further supported by the absence of a strong dependence of ionising luminosity on Galactocentric radius or spiral arm environment in Sections \ref{subsec: photon flux and distance} and \ref{subsec: HII_Galactic distribution}, and in Figure \ref{fig: ionising photos_n_multiple panels}, suggesting that local molecular cloud conditions are more important than large-scale Galactic environment in determining the strength of massive star formation.

Generally, these results demonstrate that the radio–cloud association framework not only improves the distance determinations of the sources but also provides direct insight into the physical interplay between \ion{H}{ii} regions and molecular clouds. By identifying and resolving RRL velocity ambiguities, kinematic distance, and velocity outliers, our analysis establishes a strong and solid observational basis for studying massive star formation and
feedback across the Galaxy. This paper and dataset therefore forms a valuable groundwork for future research  on cloud evolution, stellar feedback, and the Galactic scale organisation of large scale star-forming regions.

\section{Conclusions}
\label{sec: Conclusions}

In summary, this study has presented a comprehensive velocity and kinematic distance assignments of Galactic \ion{H}{ii} regions by combining WISE RRL and SMGPS radio continuum data with molecular cloud information from the SEDIGISM $^{13}$CO (2--1) survey. A total of 741 \ion{H}{ii} regions have been identified after successfully associated with 684 SEDIGISM molecular clouds and assigned CO-based systemic velocities. For analyses involving heliocentric distances and derived physical properties, we restricted the sample to the 640 \ion{H}{ii} regions associated with 572 molecular clouds having reliable kinematic distances ($d_\text{reliable}=1$). Among the 640 sources, 257 already have published kinematic distances and RRL velocity measurements, while the remaining 383 did not previously possess distance information, even though a fraction of them had velocity measurements. Thus, this work increases the number of WISE \ion{H}{ii} regions with robust and reliable kinematic distances by $\sim$60 per cent (adding 383 new distances) and provides a homogeneous recalibration for 257 previously catalogued regions.
The following are our main findings:

\begin{enumerate}[leftmargin=*, nosep]

    \item Using the emission fraction metric, \fw\,, to rank coherent molecular emission windows, we are able to assign the most physically meaningful velocity to each SMGPS source. This allows our association method to resolve cases where \ion{H}{ii} regions have multiple reported RRL velocities, identifying the molecular component most likely to represent the natal environment of the radio continuum source. The method therefore provides a systematic framework for linking \ion{H}{ii} regions to their parental molecular clouds through combined spatial and kinematic coherence. Furthermore, the lack of a systemic trend or correlation between the ionising photon flux of the radio continuum sources and their adopted reliable CO-based distances indicates that the assigned kinematic distances are robust. This lack of trend further reinforces the physical association between the radio continuum sources and their parent molecular clouds across a wide range of massive star formation environments.
    
    \item Using a validation sample of 329 \ion{H}{ii} regions with single RRL velocity measurements from the literature, we find that the CO-derived velocities show excellent agreement with independent WISE RRL measurements, with a median difference of 3.46 \kms. This confirms that the SEDIGISM molecular cloud velocities reliably trace the systemic velocities of \ion{H}{ii} regions, forming a solid foundation for deriving reliable kinematic distances using the CO-based association method. This method has enabled us to assign velocities and kinematic distances to \ion{H}{ii} regions without RRL detections. This has not only substantially increased the sample of \ion{H}{ii} regions with known velocities and distances, but also increases the sample of molecular clouds known to have HMSF (identifying a new 515 molecular clouds with active HMSF that previously lacked such HMSF tracers in the SEDIGISM catalogue).

    \item Our analysis of molecular cloud properties and CO moment maps reveals that many \ion{H}{ii} regions are embedded within, or closely associated with, massive molecular clouds that are gravitationally bound or marginally bound, and that exhibit elevated star formation efficiencies and enhanced dense gas fractions. Moreover, it is evident that the most luminous \ion{H}{ii} regions are closely linked to the massive molecular clouds. This is consistent with the paradigm that massive clouds provide the fundamental and necessary mass reservoir required for forming massive star clusters \citep{2014prpl.Dobbs, 2018MNRAS...Urquhart, 2021MNRAS...Urquhart}.
    
    \item The surrounding molecular environments exhibit velocity gradients, broadened linewidths, and complex gas structures, indicating that feedback from massive stars plays an important role in shaping cloud structure, influencing cloud evolution, and potentially regulating ongoing star formation. Furthermore, several of these associated clouds display ring-like or shell-like morphologies--which is associated with large scale star formation, however, the dominant cloud morphologies are elongated structures, consistent with the morphological classification of molecular clouds identified in recent SEDIGISM studies \citep[e.g.][]{2022A&A...Neralwar, 2022A&A...Neralwar_b}.

    \item At Galactic scales, the spatial distribution of the \ion{H}{ii} regions follows the loci of the major spiral arms traced by the SEDIGISM molecular cloud population, particularly the Scutum-Centaurus and Norma arms, reinforcing the close connection between massive star formation and spiral-arm molecular structure. However, we find no strong systematic dependence of ionising photon flux on Galactocentric radius or spiral-arm environment, with a large intrinsic scatter present throughout the sample. Additionally, we also find no enhanced star formation efficiency (SFE) effect inherently driven by the spiral arms. Instead, it is the molecular gas environments that dictate massive star formation in these regions. This suggests that local molecular cloud conditions play a more dominant role in governing massive star formation than large-scale Galactic environment alone.
    
\end{enumerate}

Our \ion{H}{ii} region distance determination method lays a good foundation for further determination of distances for other \ion{H}{ii} regions that still lack kinematic distances; and the catalogue of distances and physical associations presented here serves as a foundational resource for future studies. It enables targeted investigations of feedback physics in resolved cloud-\ion{H}{ii} region complexes, statistical studies of massive cluster formation environments, and more precise modeling of the Milky Way's star formation rate and, perhaps, chemical enrichment. The cloud association method itself is directly applicable to future synergies between next-generation radio surveys and molecular line maps, promising a fully resolved view of star formation across the Galactic disc.

\section*{Acknowledgements}

This study uses the WISE data, the SEDIGISM data acquired with the Atacama Pathfinder EXperiment (APEX) and the MeerKAT data obtained from the MeerKAT telescope. WISE is a joint project of the University of California, Los Angeles, and the Jet Propulsion Laboratory/California Institute of Technology, funded by the National Aeronautics and Space Administration (NASA). The WISE catalogue was created by Loren Anderson, Thomas Bania, Dana Balser, Virginia Cunningham, Trey Wenger, Brittany Johnstone, and William Armentrout, with help from numerous students at West Virginia University. APEX is a collaboration among the Max-Planck-Institut fur Radioastronomie, the European
Southern Observatory, and the Onsala Space Observatory. MeerKAT telescope is operated by the South African Radio Astronomy Observatory, which is a facility of the National Research Foundation, an agency of the Department of Science and Innovation. This research would not have been possible without the Astropy project and the NASA ADS.
We acknowledge and thank the Development in Africa with Radio Astronomy (DARA) project for supporting this research through the UK’s Science and
Technologies Facilities Council (STFC) grant ST/Y006100/1.  MAT acknowledges support from STFC grant awards ST/R000905/1 and
ST/W00125X/1. MOL extends his gratitude to the University of Leeds for their support during his visits.

\section*{Data Availability}

The study enjoys the use of data from:\\
The Wide-field Infrared Survey Explorer (WISE) Catalogue of Galactic \ion{H}{ii} Regions \citep[][]{2014ApJS..Anderson, 2015ApJS..Anderson, 2018ApJS..Anderson}, that are available at \url{http://astro.phys.wvu.edu/wise}.\\
The SMGPS survey data \citep{2024MNRAS..Goedhart} and the SMGPS Extended Source Catalogue \citep{2025A&A...Bordiu}, that are available at \url{https://doi.org/10.48479/3wfd-e270}, and \url{https://doi.org/10.48479/t1ya-na33}, respectively. The Extended Source Catalogue is also available at the CDS via anonymous ftp to \url{https://cdsarc.cds.unistra.fr/viz-bin/cat/J/A+A/695/A144} or via \url{http://cdsweb.u-strasbg.fr/cgi-bin/qcat?J/A+A/}.\\
The SEDIGISM survey, constructed by James Urquhart and hosted by the Max Planck Institute for Radio Astronomy, which includes projects 092.F-9315 and 193.C-0584, and the processed data products are available at \url{https://sedigism.mpifr-bonn.mpg.de/index.html}.\\



\bibliographystyle{mnras}
\bibliography{paper} 



\appendix

\section{SMGPS-SEDIGISM-WISE merged catalogue table format}
\label{appendix_A}

\begin{table*}
\centering
\caption{Format of the SMGPS-SEDIGISM-WISE merged catalogue table for the associated clouds and radio continuum sources.}

\begin{tabular}{c|c|p{12cm}}
\hline
\hline
Name & Unit & Description \\
\hline
iauName & - & SMGPS extended source name in IAU format (from {\citealt{2025A&A...Bordiu}}) \\
$\ell_s$ & degree & SMGPS source centroid position: Galactic Longitude coordinate (from {\citealt{2025A&A...Bordiu}}) \\
$b_s$ & degree & SMGPS source centroid position: Galactic Latitude coordinate (from {\citealt{2025A&A...Bordiu}}) \\
radius & - & SMGPS source radius in pixels (from {\citealt{2025A&A...Bordiu}}) \\
radius\_wcs & arcmin & SMGPS source angular size (radius) (from {\citealt{2025A&A...Bordiu}}) \\
flux & Jy & SMGPS source measured flux density with background and nested compact sources subtracted (from {\citealt{2025A&A...Bordiu}})
\\
classname & - & SMGPS source classification label in string format (from {\citealt{2025A&A...Bordiu}}) \\
classid & - & SMGPS source classification id e.g. 0=Unknown, 2=Galaxy, 3=PN, 4=SNR, 5=Bubble, 6=\ion{H}{ii} (from {\citealt{2025A&A...Bordiu}}) \\
objname & - & SMGPS source object name (from {\citealt{2025A&A...Bordiu}}). It is also the matching object name which is the WISE name (from {\citealt{2014ApJS..Anderson,2015ApJS..Anderson, 2018ApJS..Anderson}}), in case of one-to-one associations \\
cloud\_name & - & SEDIGISM molecular cloud name as per the SEDIGISM naming scheme (from {\citealt{2021MNRAS.Duarte}}) \\
$\ell_c$ & degree & SEDIGISM cloud centroid position in Galactic longitude coordinate (from {\citealt{2021MNRAS.Duarte}}) \\
$b_c$ & degree & SEDIGISM cloud centroid position in Galactic latitude coordinate (from {\citealt{2021MNRAS.Duarte}}) \\
cloud\_$v_\text{lsr}$ & \kms & SEDIGISM cloud systemic and centroid velocity (from {\citealt{2021MNRAS.Duarte}}) \\
Mean SDG\_$v_\text{lsr}$ & \kms & Mean systemic velocity from velocities of all SEDIGISM clouds intersecting the best-matching velocity window (from {\citealt{2026MNRAS...Langa}}) \\
velocity\_range & \kms & Window velocity range for the best-matching velocity window where all the associated clouds fall (from {\citealt{2026MNRAS...Langa}})\\
linewidth ($\sigma_\text{v}$) & \kms & SEDIGISM cloud velocity dispersion (from {\citealt{2021MNRAS.Duarte}}) \\
HMSF & - & Cloud High mass star formation (from {\citealt{2021MNRAS.Duarte}}) \\
cloud\_dist\_kpc & kiloparsec (kpc) & Cloud final adopted physical distance (from {\citealt{2021MNRAS.Duarte}}, updated by {\citealt{2022A&A...Colombo}}) \\
cloud\_mean\_dist\_kpc & kiloparsec (kpc) & mean cloud final adopted physical distance from distances of all SEDIGISM clouds intersecting the best-matching velocity window (updated by {\citealt{2026MNRAS...Langa}})\\
d\_flag & - & Flag describing the method by which the cloud's final distance was decided e.g. -1=no distance, 0=maser, 1=no ambiguity, 2=tangent, 3=dark cloud, 4=IRDC, 5=lit.HISA, 6=dir.HISA, 7=ATLASGAL near, 8=Solomon, 9=size-linewidth, 10=ATLASGAL far, 11=extinction, 12=ambiguous, 13=spiral arms (from {\citealt{2021MNRAS.Duarte}}, updated by {\citealt{2022A&A...Colombo}}) \\
d\_solution & - & Cloud flag describing the type of kinematic distance solution, e.g. NA=not ambiguous, T=tangent, N=near, F=far, M=maser (updated by {\citealt{2021MNRAS.Duarte}}) \\
d\_reliable & - & Cloud flag to indicate sources with a reliable distance, e.g. 1=reliable, 0=non-reliable; (from {\citealt{2021MNRAS.Duarte}}, updated by {\citealt{2022A&A...Colombo}}) \\
Mass & solMass (M$_\odot$) & SEDIGISM cloud mass (from {\citealt{2021MNRAS.Duarte}}, updated by {\citealt{2022A&A...Colombo}}) \\
R\_dec & parsec (pc) & Cloud deconvolved equivalent radius (from {\citealt{2021MNRAS.Duarte}}, updated by {\citealt{2022A&A...Colombo}}) \\
column\_density ($N_{\mathrm{H}_2}$) & \pcmm & SEDIGISM cloud average column density (from {\citealt{2021MNRAS.Duarte}}, updated by {\citealt{2022A&A...Colombo}}) \\
surf\_density ($\Sigma$) & M$_\odot$\pc & Cloud deconvolved average gas surface density (from {\citealt{2021MNRAS.Duarte}}, updated by {\citealt{2022A&A...Colombo}}) \\
$\alpha_\mathrm{vir}$ & - & Cloud deconvolved virial parameter (from {\citealt{2021MNRAS.Duarte}}, updated by {\citealt{2022A&A...Colombo}}) \\
SFE & L$_\odot$/M$_\odot$ & Cloud star formation efficiency (from {\citealt{2021MNRAS...Urquhart}}) \\
DGF & - & Cloud dense gas fraction (from {\citealt{2021MNRAS...Urquhart}}) \\
emission\_fraction (\fw) & - & The ratio of the total integrated intensity within the best emission window to the total integrated intensity of all emission windows in an extracted spectrum (from {\citealt{2026MNRAS...Langa}})\\
WISE\_name & - & \ion{H}{ii} region name as per the WISE catalogue naming scheme (from {\citealt{2014ApJS..Anderson, 2015ApJS..Anderson, 2018ApJS..Anderson}}) \\
$\ell_w$ & degree & WISE \ion{H}{ii} region centroid position in Galactic longitude coordinate (from {\citealt{2014ApJS..Anderson, 2015ApJS..Anderson, 2018ApJS..Anderson}}) \\
$b_w$ & degree & WISE \ion{H}{ii} region centroid position in Galactic latitude coordinate (from {\citealt{2014ApJS..Anderson, 2015ApJS..Anderson, 2018ApJS..Anderson}}) \\
WISE\_$v_\text{lsr}$ & \kms & Assigned systemic velocity to the WISE \ion{H}{ii} regions (from {\citealt{2014ApJS..Anderson, 2015ApJS..Anderson, 2018ApJS..Anderson}}) \\
WISE\_$v_\text{lsr}$\_reference & - & Different authors that assigned WISE \ion{H}{ii} region velocities (from {\citealt{2014ApJS..Anderson, 2015ApJS..Anderson, 2018ApJS..Anderson}}) \\
Adopted SMGPS\_$\text{SDG}$ vel & \kms & CO-derived systemic velocity assigned to the SMGPS \ion{H}{ii} region as a result of the SEDIGISM cloud association with SMGPS \ion{H}{ii} region (from {\citealt{2021MNRAS.Duarte}}) \\
SMGPS\_$\text{SDG}$ vel\_reference & - & Author of the CO-based centroid velocity assigned to the SMGPS \ion{H}{ii} region (from {\citealt{2021MNRAS.Duarte}}) \\
WISE\_dist\_kpc & kiloparsec (kpc) & WISE \ion{H}{ii} region distance assignments (from {\citealt{2014ApJS..Anderson, 2015ApJS..Anderson,2018ApJS..Anderson}}) \\
WISE\_dist\_Method & - & Method employed to assign WISE \ion{H}{ii} region distances (from {\citealt{2014ApJS..Anderson, 2015ApJS..Anderson, 2018ApJS..Anderson}}) \\
WISE\_dist\_reference & - & List of authors involved in assigning WISE \ion{H}{ii} region distances (from {\citealt{2014ApJS..Anderson, 2015ApJS..Anderson, 2018ApJS..Anderson}}) \\
source\_adopted\_dist\_kpc & kiloparsec (kpc) & \ion{H}{ii} region distance adopted based on the cloud association method (from {\citealt{2021MNRAS.Duarte}}, updated by {\citealt{2022A&A...Colombo}}) \\
source\_adopted\_dist\_Method & - & Cloud association method used to assign distances to the SMGPS extended \ion{H}{ii} region \\
source\_adopted\_dist\_reference & - & Author of the SEDIGISM cloud distances assigned and adopted by the SMGPS extended \ion{H}{ii} regions (from {\citealt{2021MNRAS.Duarte}}, updated by {\citealt{2022A&A...Colombo}}) \\

\hline
\end{tabular}

\label{tab: SMGPS_SED_WISE_merged_catalogue}
\end{table*}


\bsp	
\label{lastpage}
\end{document}